\documentclass[twocolumn, usenatbib]{mnras}
\usepackage{savesym}
\usepackage{graphicx}
\usepackage{longtable}
\usepackage{changepage}

\expandafter\let\csname equation*\endcsname\relax
  \expandafter\let\csname endequation*\endcsname\relax 
\usepackage{subfig}
\usepackage{amsmath}
\usepackage{amssymb}
\usepackage{verbatim}
\usepackage[yyyymmdd,hhmmss]{datetime}
\usepackage{array}
\usepackage{times}
\usepackage{xcolor}
\usepackage{xspace}

\DeclareRobustCommand{\DE}[3]{#2}
\let\DEthebibliography\thebibliography
\def\thebibliography{\DeclareRobustCommand{\DE}[3]{##3}\DEthebibliography}

\newcommand{\beq}{\begin{equation}}
\newcommand{\eeq}{\end{equation}}

\newcommand{\sk}{{\tt superkerr}\xspace}
\newcommand{\skl}{{\tt skline}\xspace}
\newcommand{\skc}{{\tt skconv}\xspace}

\title[Super-extremal Iron lines]{ Reflecting on naked singularities: iron line fitting as a  probe of the cosmic censorship conjecture  }
\author [Andrew Mummery, Adam Ingram]{Andrew Mummery$^1$\thanks{E-mail:
andrew.mummery@physics.ox.ac.uk}, Adam Ingram$^2$
\\
$^1$Oxford Theoretical Physics, Beecroft Building,  Clarendon Laboratory, Parks Road, Oxford, OX1 3PU, United Kingdom\\
$^2$School of Mathematics, Statistics and Physics, Newcastle University, Herschel Building, Newcastle upon Tyne, NE1 7RU, UK }

\date{}

\begin{document}

\pagerange{\pageref{firstpage}--\pageref{lastpage}} \pubyear{2023}

\maketitle

\label{firstpage}

\begin{abstract} 
We demonstrate that the X-ray iron line fitting technique can be leveraged as a powerful probe of the cosmic censorship conjecture.  We do this by extending existing emission line models to arbitrary spin parameters of the Kerr metric, no longer restricted to black hole metrics with $|a_\bullet |< 1$.  We show that the emission lines from naked singularity metrics ($|a_\bullet| > 1$) show significant differences to their black hole counterparts, even for those metrics with identical locations of the innermost stable circular orbit, i.e., emission line fitting does not suffer from the degeneracy which affects continuum fitting approaches. These differences are entirely attributable to the disappearance of the event horizon for $|a_\bullet| > 1$. We highlight some novel emission line features  of naked singularity metrics, such as ``inverted'' emission lines (with sharp red wings and extended blue wings) and ``triple lines''. The lack of detection of any of these novel features provides support of the cosmic censorship conjecture.   We publicly release {\tt XSPEC} packages \skl and \skc which can now be used to probe the cosmic censorship conjecture in Galactic X-ray binaries and Active Galactic Nuclei. The inclusion of super-extremal spacetimes can be alternatively posed as a way of stress-testing conventional models of accretion. 
\end{abstract}

\begin{keywords}
accretion, accretion discs --- black hole physics 
\end{keywords}
\noindent
%Complied at \today\ \currenttime\ .

\section{Introduction}

% Reviews: \citep{Reynolds2021,Bambi2021}

Asymmetric and broadened iron emission lines are observed in the X-ray spectrum of accreting compact objects across the entire mass scale \citep[e.g.][]{Miller2007}. This includes supermassive accretors in Active Galactic Nuclei \citep{Tanaka1995,Fabian2000} and stellar-mass accretors in Galactic X-ray binary systems \citep{Miller2015}, including neutron star systems \citep{Cackett2008,Ludlam2019}. The line results from the irradiation of the accretion disk by hard X-rays from hot plasma located close to the central object. This process is typically referred to as `reflection', although it is more accurately described as reprocessing and re-emission \citep{Bambi2021}. The iron K$\alpha$ fluorescence line at $\sim 6.4$ keV is a prominent feature of the reflection spectrum, which also includes emission lines and absorption edges from all astrophysically abundant elements and a broad Compton scattering feature at $\sim 20-30$ keV known as the `Compton hump' \citep{Matt1991,Ross2005,Garcia2013}. The iron line is a particularly powerful diagnostic because it is narrow in the emission rest frame, whereas the observed line profile is heavily distorted by a combination of the relativistic orbital motion of the emitting material, and the gravitational energy-shifting of the emitted photons over their trajectory to the observer \citep{Fabian1989,Laor1991,Dauser2010}. 

The line profile is sensitive to the location of the inner edge of the accretion disc, as this sets the fastest orbital frequency and the largest gravitational redshift. As such, it provides a powerful probe of the dimensionless spin parameter of the Kerr metric, $a_\bullet$, under the assumption that the disc inner edge is set by the innermost stable circular orbit (ISCO), which is itself a function of spin. The spin parameter also influences the line profile via the spin-dependence of orbital frequency, gravitational redshift, and photon trajectories, although these are more subtle effects. As such, 
analysis of observed broad iron emission lines has been used to place constraints on numerous Kerr metric spin parameters in the literature, under the physical assumption that these objects are black holes \citep{Middleton2016,Reynolds2021}. The black hole assumption amounts to restricting the spin parameter to values $|a_\bullet| < 1$. The Kerr metric, however, remains a mathematically valid solution of Einstein's field equations for all spin parameters $a_\bullet$, including metrics with $|a_\bullet| > 1$, which instead describe the spacetime external to a naked singularity. 

While remaining a valid solution in General Relativity, the conventional  assumption is that these naked singularity solutions will not be present in nature.  This assumption (more generally that all solutions of Einstein's equations which occur in nature and which host singularities have those singularities hidden behind an event horizon) is known as the Weak Cosmic Censorship Conjecture \citep{Penrose1969}, and while unproven does have supporting evidence \citep{Wald97}. Principally this supporting evidence comes from theoretical modelling of the gravitational collapse and formation of singularities, but it seems likely that X-ray spectral information (which probes the spacetime close to the singularity) has much to offer the study of this conjecture.  

{The properties of super-spinning Kerr metrics have received renewed interest in recent years, owing to the realisation that string theory may allow for an external spacetime described by a super-extremal Kerr metric, which is grafted onto an interior string-theory metric which shrouds the singularity \citep[see][for a detailed discussion]{Gimon09}. While no exact interior solution in three spatial (and one temporal) dimensions is  known, a four space (one time) dimension interior solution with these properties is known \citep{Gimon04}, suggesting that such a solution may exist for three spatial dimensions. If such a string theory ``resolution'' to the naked singularities of super-spinning Kerr metrics exists, then it remains possible that these objects may exist in our Universe without breaking the Cosmic Censorship Conjecture, and can in principle be probed by X-ray observations. While an interior string theory solution is expected to shroud the singularity (and as such breaks the strict association between $|a_\bullet| > 1$ and the cosmic censorship hypothesis), it is not as of yet clear on what spatial scale $R$ this shrouding will occur. In this paper we assume $R\ll r_g$, where the gravitational radius $r_g \equiv GM/c^2$, and do not curtail the Kerr metric at any finite radius.   }

{Inspired by this string theory argument \cite{Stuchlik10} computed photon energy-shifts for observers in the super-extremal Kerr metric, before extending this work in \cite{Schee13} to examine some emission line properties of super-spinning Kerr singularities, focusing on line profiles produced from a disc with uniform emissivity. Similarly, \cite{Ranea-Sandoval15} studied the properties of both black hole and super-extremal iron lines, but focused on the effects which strong magnetic fields have on the properties of these lines. In this work we consider more general treatments of super-extremal emission lines, both by examining a wider range of spins, but also more realistic disc rest-frame emissivity profiles. } 

{Other related work involves iron line calculations in non-Kerr metrics, which are in some ways similar to the Kerr metric. These metrics are typically either solutions of modified theories of gravity, or which include additional beyond the standard model physics.  \cite{Liu18} examine iron line profiles about the naked singularity  Janis-Newman-Winicour metric (which assumes that a charged and massless scalar field $\Phi$ is present throughout space). While \cite{Zhou20} compute iron line  profiles in a modified gravity version of the Kerr metric, where black hole metrics (i.e., metrics with an event horizon) can be constructed with spin parameters $|a_\bullet | > 1$. These modified gravity metrics have additional free parameters which allow event horizons to be present in metrics with $a_\bullet > 1$.  The philosophy of these papers differs somewhat from our approach, in that these analyses seek to probe whether General Relativity is the correct theory of gravity, while we are seeking to probe whether accreting compact objects have event horizons.     }

As far as the authors are aware, no existing publicly available  relativistic reflection models \citep[e.g.][]{Laor1991,Dauser2013} can be used in the super-extremal ($|a_\bullet| > 1$) limit of the Kerr metric \citep[although some publicly available models are based on modified gravity metrics in place of the Kerr metric e.g.,][]{Bambi2017,relxillnkref}.
As such, existing models must be extended, which is the purpose of this paper.  

This is not just an extension of theoretical interest however. The need for this extension can be motivated entirely from a data-driven perspective, since measurements preferring very rapid spin parameters can run into the prior upper bound imposed by the black hole assumption. This often occurs in studies of both Active Galactic Nuclei \citep[e.g.][]{Vasudevan2016} and X-ray binaries \citep[e.g.][]{Draghis2023}. For example, a recent analysis of the X-ray spectrum of the X-ray binary \mbox{Cyg X-1} (which contained an iron line reflection component, in addition to a thermal continuum) found an extremely rapid rotation rate with a 3$\sigma$ lower bound of $a_\bullet > 0.9985$, and a best fit limited by the prior of $a_\bullet \leq 0.9999$ \citep{Zhao21}. \mbox{Cyg-X1} has repeatedly been found to have a high spin from numerous iron line studies \citep[e.g.,][]{Fabian12, Tomsick14, Duro16, Walton16}. It is of critical importance that such inferences of high spins are robust to the removal of the spin prior placed on conventional models. This can only be tested with the extended model we present here. In addition to probing the cosmic censorship conjecture, the inclusion of super-extremal Kerr spacetimes in fitting procedures can equally well be thought of as stress-testing conventional models of accretion.  

{In this paper, we focus principally on the properties of line profiles resulting from a $\delta$-function fluorescence line being emitted in the plasma rest-frame. This enables us to examine the observable properties that the event horizon (or lack thereof) imprints onto a single emission line. We release this line profile model publicly as the {\tt XSPEC} model \skl. We additionally define a convolution model \skc, that can be convolved with a model of the full rest-frame reflection spectrum containing numerous lines, edges and scattering features to provide a much more complete description of relativistic reflection.}
% \cmtmum{In this paper we focus principally on the individual line profiles resulting from a $\delta$-function photon emissivity in the disc rest frame (i.e., all emitted photons have the same rest-frame frequency). The purpose of this is to examine the observed properties which the event horizon (or lack thereof) imprint onto a single emission line. As mentioned above, a reflection spectrum consists of a large number of both continuum and line features, rather than just a single line. However, any rest frame emission profile can be described as a convolution with the $\delta$-function profiles computed here, and so this procedure elucidates the key physics of naked singularity reflection spectroscopy. For future modelling \skl can be included as a convolution model in {\tt XSPEC}, and so can be extended to more general rest frame emissivities.   }
Coupled with our recent extension of disc-continuum fitting models to the super-extremal regime \citep{MummeryBalbusIngram24}, all conventional X-ray spectral modelling techniques can now be performed in the extended Kerr metric.  

The layout of this paper is the following. In section \ref{full-kerr-metric} we recap the properties of the Kerr metric in both the black hole and naked singularity regimes. In section \ref{line-fitting-formalism} we introduce the X-ray line fitting formalism, and describe the fundamental equations which we solve. In section \ref{image-plane-maps} we present an analysis of the structure of black hole and naked singularity spacetimes from the perspective of observed photon energy shifts, the relevant quantity for emission line analysis. In section \ref{line-profiles} we present our analysis of emission line profiles in naked singularity metrics, highlighting their potential for probing the cosmic censorship conjecture. In section \ref{conv} we show how the peculiar features of iron lines observed in naked singularity metrics are observable when a more complex rest frame spectrum is taken into account, before concluding in section \ref{conclusions}.

\section{The general Kerr spacetime }\label{full-kerr-metric}
A detailed discussion of the properties of the super-extremal Kerr metric was presented in \cite{MummeryBalbusIngram24}, to which we refer the reader for full details, providing only a  summary of the key properties here.    Our notation is as follows.   We use physical units in which we denote   the speed of light by  $c$, and the Newtonian gravitational constant by  $G$.    In coorindates $x^\mu$, the invariant line element ${\rm d}\tau$ is given by
\beq
{\rm d}\tau^2 = - g_{\mu\nu} {\rm d}x^\mu {\rm d}x^\nu,
\eeq
where $g_{\mu\nu}$ is the metric tensor with spacetime indices $\mu, \nu$.   The coordinates are standard $(t, r, \phi, \theta)$ Boyer-Lindquist, where $t$ is time as measured at infinity and the other symbols have their usual quasi-spherical interpretation.    We shall assume that the disc exists exclusively in the Kerr midplane $\theta =\pi/2$.   For singularity  mass $M$, and angular momentum parameter $a$ (which has dimensions of length and is {\it unrestricted} in this work), the line element is:
\begin{multline}\label{metric}
    {\rm d}s^2 = - \left(1 - {2r_g r \over r^2 + a^2 \cos^2\theta }\right) c^2 {\rm d}t^2 - {4 r_g r a c \sin^2 \theta \over r^2 + a^2 \cos^2\theta} \, {\rm d}t \, {\rm d}\phi \\ + {r^2 - 2r_gr + a^2 \over r^2 + a^2 \cos^2\theta }\, {\rm d}r^2 + (r^2 + a^2 \cos^2 \theta ) \, {\rm d}\theta^2  \\ + \left(r^2 + a^2 + {2 r_g a^2 r \sin^2 \theta \over r^2 + a^2 \cos^2 \theta } \right)\sin^2 \theta \, {\rm d}\phi^2 .
\end{multline}
  Note that our choice of coordinates has asymptotic (large radius) metric signature $(-1, +1, +1, +1)$. 

\subsection{(A lack of an) Event horizon }
The outer event horizon of the Kerr metric $r_+$ is defined as the larger of the two radii at which 
\beq
{1 \over g_{rr}(r_+)} = 0 \to r_+^2 - 2r_g r_+ + a^2 = 0. 
\eeq
The solution of this condition is trivial 
\beq
r_+ = r_g + \sqrt{r_g^2 - a^2} ,
\eeq
but highlights a crucial point of physics, namely that if 
\beq
|a| > r_g ,
\eeq
then the Kerr metric has no event horizon. This of course does not mean that the metric  ceases to be a solution of the governing equations of general relativity, merely that the singularity at $r=0$ is not ``hidden'' behind an event horizon.  As discussed in the introduction, this ``naked singularity" behaviour is hypothesised to be forbidden in astrophysically realistic settings \citep{Penrose1969}.

\subsection{Orbital motion and super-extremal ISCOs }
\begin{figure}
\centering
\includegraphics[width=0.5\textwidth]{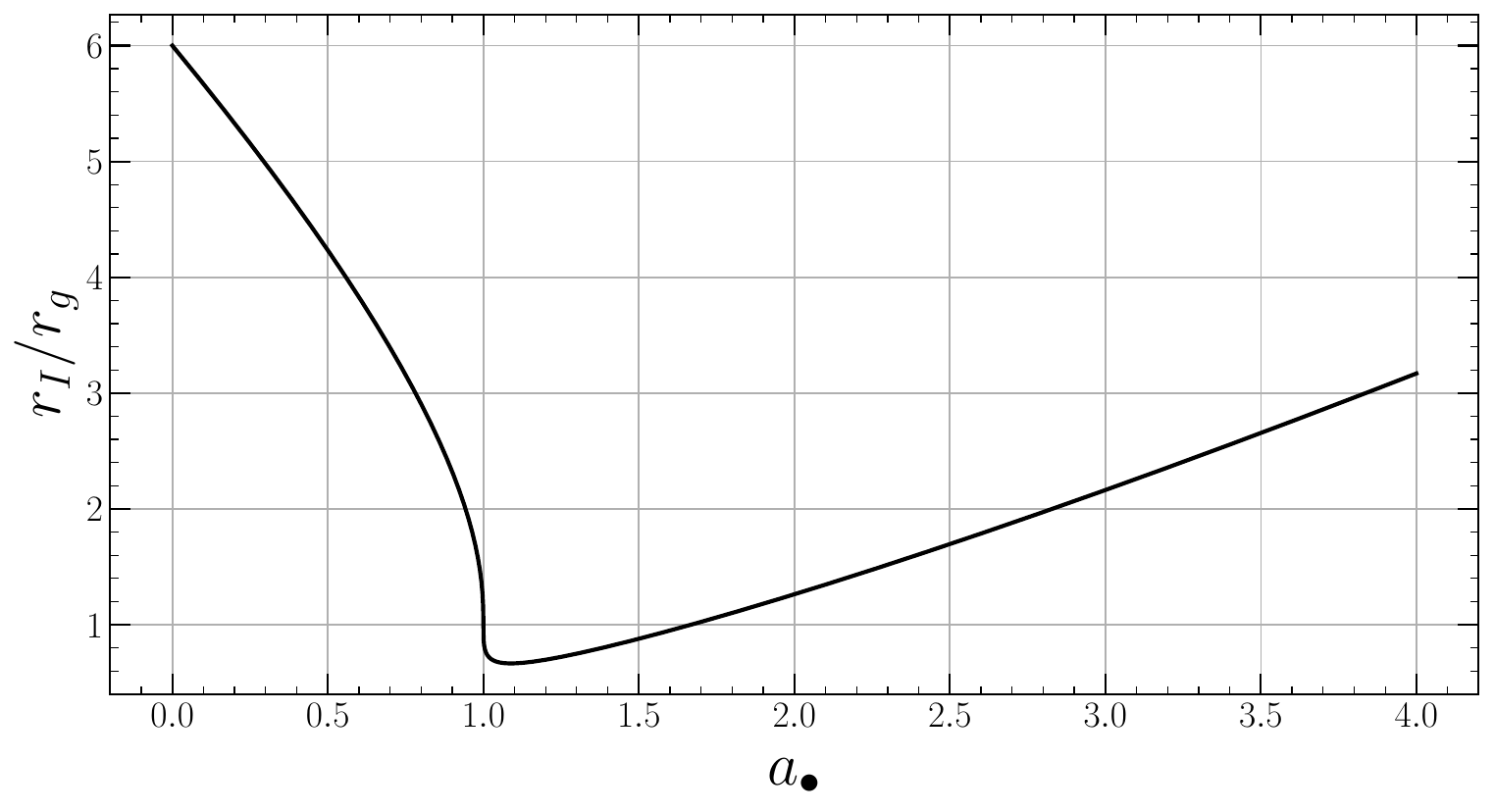}
\caption{The ISCO as a function of Kerr metric spin parameter. The ISCO exists for all spin parameters, and is a double valued function of $a_\bullet$.  }
\label{isco-loc}
\end{figure}

The key orbital component relevant for the study of accretion flows is the angular momentum of a fluid element undergoing circular motion, as this is an excellent approximation to the dynamical behaviour of the fluid in the stable disc regions. For the Kerr metric standard methods \citep[e.g.,][]{Hobson06} lead to the following circular angular momentum profile  
\beq\label{orbang}
U_\phi = (GMr)^{1/2}{(1+a^2/r^2-2ar_g^{1/2}/r^{3/2}) \over \sqrt{ 1 -3r_g/r +2ar_g^{1/2}/r^{3/2}}} .
\eeq
This orbital solution is valid in both the sub ($|a| < r_g$) and super ($|a| > r_g$) extremal regimes. 
An accretion flow will persist in its classical form wherever these orbits are stable. The stability of circular orbits  requires $\partial_r U_\phi > 0$, the relativistic analogue of the Rayleigh fluid-stability criterion.   Evaluating this gradient, and setting $\partial_r U_\phi = 0$, results in the following expression for the location of the innermost stable circular orbit \citep[ISCO, e.g.][]{Bardeen72} 
\begin{equation}\label{rI_a}
r_I^2  -    6r_gr_I + 8a\sqrt{r_gr_I} - 3a^2  = 0.
\end{equation}
All spin parameters of the super-extremal Kerr metric have an ISCO, with the ISCO  radius given by \citep[cf.][]{Bardeen72} 
\beq
{r_I / r_g}  = 3 + Z_2  - {\rm sgn}(a_\bullet) \sqrt{(3 - Z_1)(3 + Z_1 + 2 Z_2)},
\eeq
where $a_\bullet \equiv a / r_g$ is the dimensionless spin, and we have defined 
\beq
Z_1 = 
\begin{cases}
1 + (1-a_\bullet^2)^{1\over3}  \left((1+a_\bullet)^{1\over3} + (1-a_\bullet)^{1\over3}\right) , \quad  |a_\bullet| \leq 1 \\
%\\
1 - (a_\bullet^2-1)^{1\over3} \left((1+|a_\bullet|)^{1\over3} - (|a_\bullet|-1)^{1\over3}\right) , \quad |a_\bullet| > 1,
\end{cases}
\eeq
and $Z_2 = \sqrt{3a_\bullet^2 + Z_1^2}$ . The ISCO location plotted as a function of Kerr metric spin is displayed in Fig. \ref{isco-loc}. 

It will be important to note for future sections that the ISCO is a double valued function of Kerr metric spin parameter $a_\bullet$.  This was shown in  \citep{MummeryBalbusIngram24} to lead to degeneracy in the spin parameter inferred from the continuum fitting method between black hole and naked singularity solutions that share a common ISCO.  In this paper we shall demonstrate that the same degeneracy does not plague the emission line fitting technique.

\section{The X-ray line fitting formalism}\label{line-fitting-formalism}

\begin{figure*}
    \centering
    \includegraphics[width=0.49\textwidth]{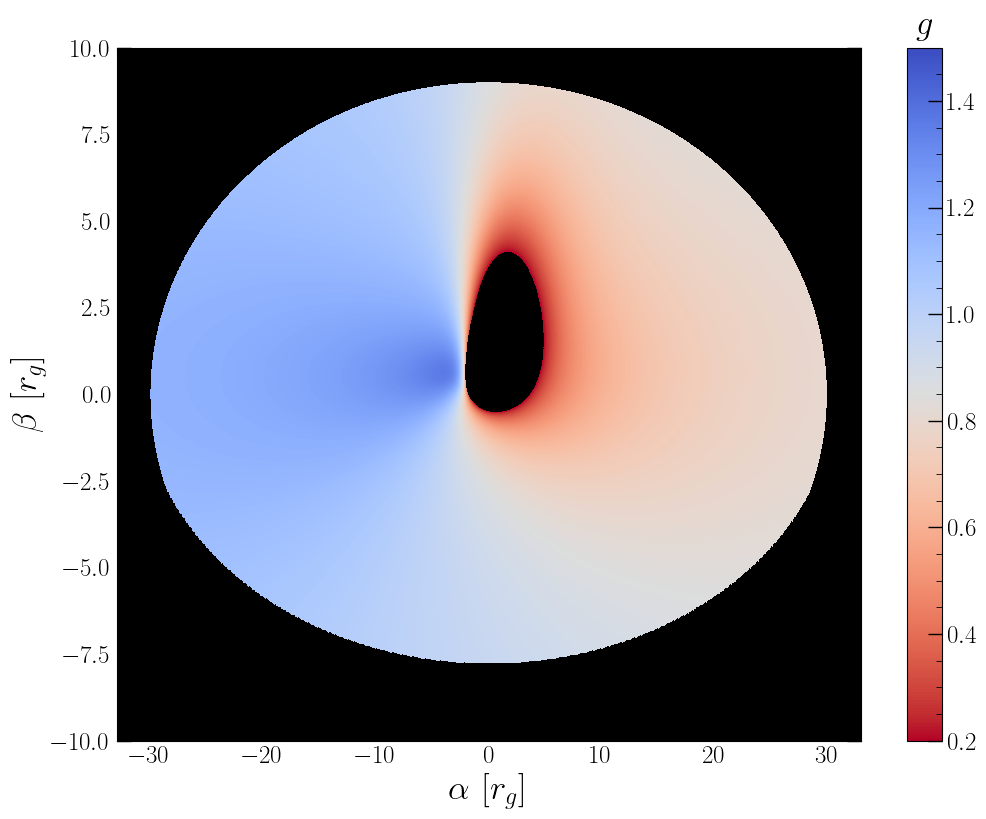}
    \includegraphics[width=0.49\textwidth]{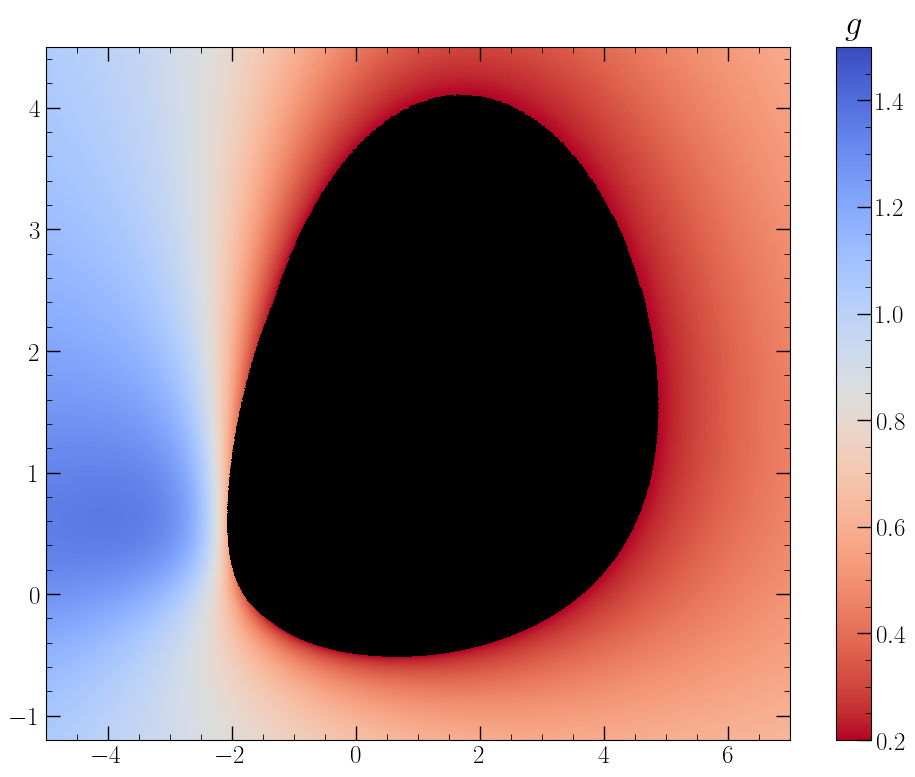}
    \includegraphics[width=0.49\textwidth]{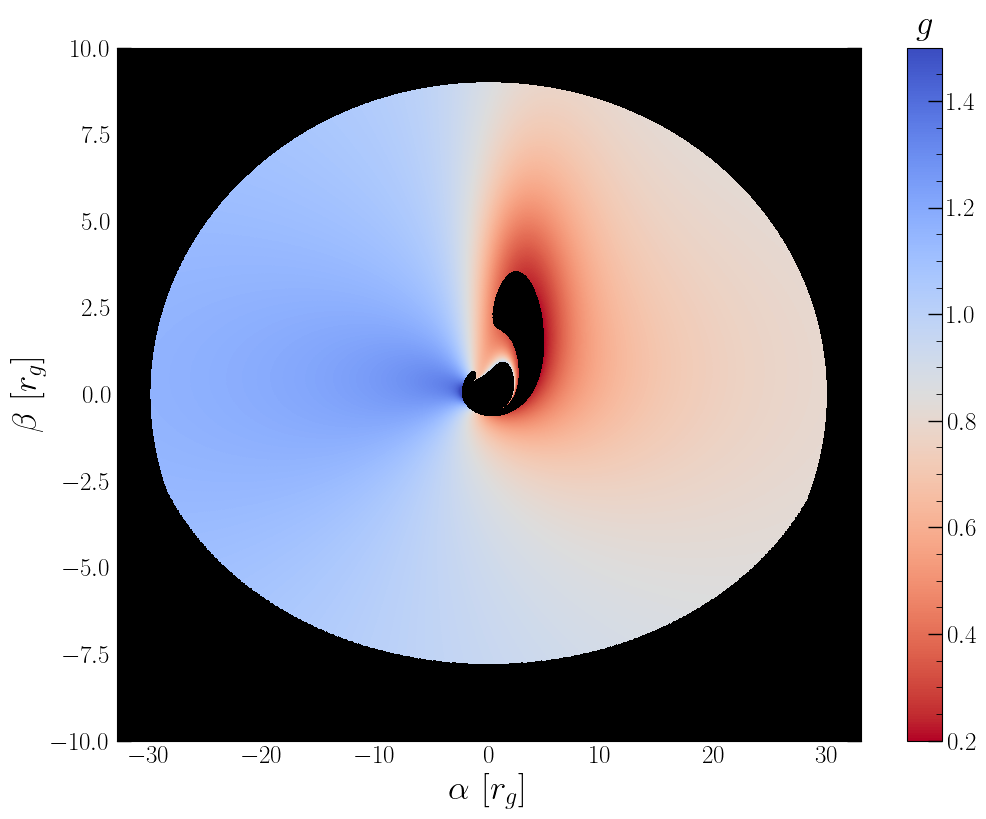}
    \includegraphics[width=0.49\textwidth]{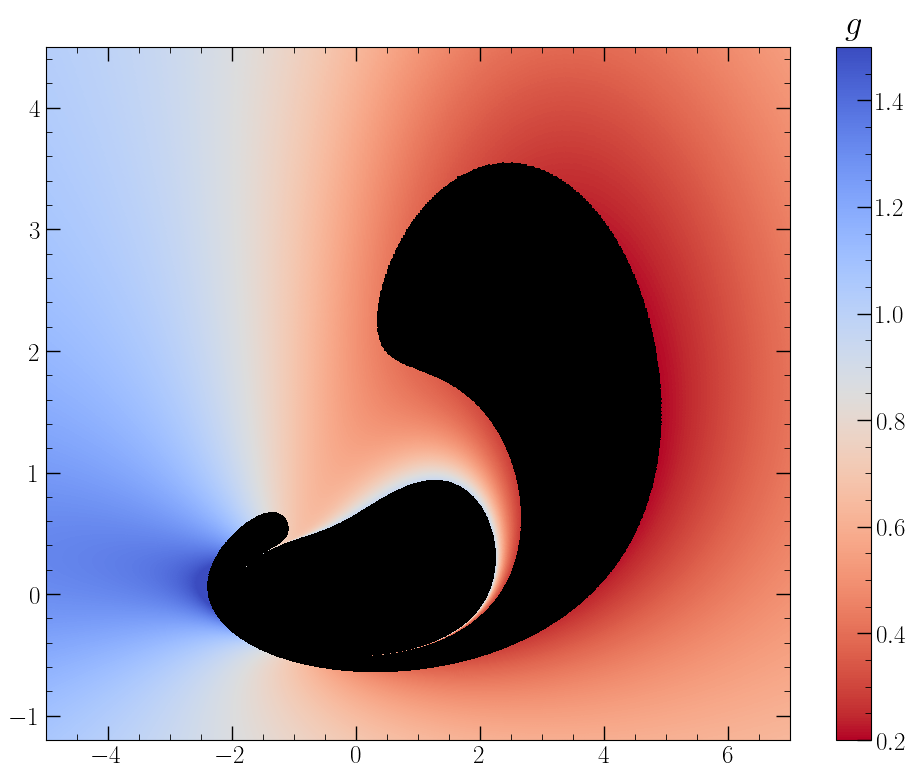}
    \caption{The disc-observer photon energy shifts of two Kerr metrics, the upper row describing a black hole with $a_\bullet = 0.998$, the lower row a naked singularity with $a_\bullet = 1.968$ (both energy shift colour bars are normalised to show the same scale). Both of these metrics have ISCO radii at the same location $r_I/r_g \simeq 1.237$. On the left we show the innermost 30 $r_g$ of both discs, while on the right only the very innermost region of the image plane is displayed, where differences are maximal.  Both systems are observed at  $\theta_{\rm obs} = 75^\circ$. }
    \label{fig:different-images}
\end{figure*}

The specific flux density $F_E$ from a source, as observed by a distant observer at rest (subscript ${\rm obs}$), is given by
\beq
F_{E}(E_{\rm obs}) = \iint I_E(E_{\rm obs} ) \, \text{d}\Omega_{{{\rm obs} }} .
\eeq 
Here, $E_{\rm obs} $ is the observed photon energy and $I_E(E_{\rm obs} )$ the specific intensity,  both measured at the location of the distant observer.   The differential element of solid angle subtended on the observer's sky by a disc area element is $\text{d}\Omega_{{{\rm obs} }}$. 
Since $I_E/ E^3$ is a relativistic invariant \citep[e.g.,][]{MTW}, we may write
\beq
F_{E}(E_{\rm obs} ) = \iint g^3 I_E(E_{\rm emit}) \, \text{d}\Omega_{{{\rm obs} }},
\eeq 
where the photon energy ratio factor $g$ is the ratio of $E_{\rm obs} $ to the emitted local rest frame photon energy $E_{\rm emit}$:
\begin{equation}\label{redshift}
g(r,\phi) \equiv 
{E_{\rm obs} \over E_{\rm emit}} = {-\left(U^\mu p_\mu\right)_{\rm obs} \over -\left(U^\nu p_\nu\right)_{\rm disc}  } = {1 \over U^t} \left(1 + {p_\phi \over p_t} {U^\phi \over U^t} \right)^{-1} . 
\end{equation}
In this final expression $p_\phi$ and $-p_t$ are the angular momentum and energy of the emitted photon (conveniently constants of motion in the Kerr metric), $U^t$ and $U^\phi$ are the time and azimuthal 4-velocity components of the orbiting disc fluid{, which are given by } 
\begin{align}
    U^t &= {1 + a\sqrt{r_g/r^3} \over \left(1 - 3r_g/r + 2a\sqrt{r_g/r^3}\right)^{1/2}}, \\
    U^\phi &=  {\sqrt{GM/r^3}  \over \left(1 - 3r_g/r + 2a\sqrt{r_g/r^3}\right)^{1/2}} . 
\end{align}
For
% an
{a stationary} observer at a large distance $D$ from the source, the differential solid angle into which the radiation is emitted is
\beq
 \text{d}\Omega_{{{\rm obs} }} = \frac{\text{d}\alpha \, \text{d} \beta}{D^2} ,
\eeq 
where $\alpha$ and $\beta$ are (Cartesian) photon impact parameters at infinity \citep[e.g.,][]{Li05}. Therefore, determining the observed specific flux from an astrophysical source breaks down into knowing the specific intensity of the source in its own rest frame, and calculating the energy shifts of the photons on their path to the observer. This second step must in general be done numerically. 

{In the simplified case whereby the emergent spectrum is assumed to be only a $\delta-$function emission line, the rest-frame specific intensity can be expressed as \citep[e.g.][]{Dauser2010}}
% Consider an accretion flow irradiated by some external source of X-ray photons (perhaps the base of a jet, or a ``corona''). If this accretion flow contains atoms which can undergo energy level transitions then by absorbing incoming X-ray photons these atoms may become excited, and upon de-excitation these atoms will emit photons with characteristic spectral line energies in their own rest frame. Iron is one such important element, and the {\textup Fe K}$\alpha$ line at 6.4 keV has been the focus of a huge amount of astronomical research. Let us assume that in its own rest frame a de-exciting atom emits a delta function spectral line\footnote{While naturally a simplification, any general rest-frame line profile $I(E)$ can be expressed as the integral $I(E) = \int I(E') \delta(E' - E) \, {\rm d}E'$. This energy integral can then be performed after the calculations in this paper, meaning that our ``Green's function'' approach contains all  the relevant physics. }, an approach put forward in \cite{Dauser10} 
\begin{equation}
    I_E(E_{\rm emit}) = I_{\rm line} \,  \delta(E_{\rm emit} - E_{\rm line}) \, \epsilon(r, \phi),
\end{equation}
{where $\epsilon(r, \phi)$ the the disc emissivity function. In this case,}
% then
the observed specific flux is
\beq\label{integral}
F_{E}(E_{\rm obs} ) = {I_{\rm line} \over D^2} \iint g^3 \epsilon(r, \phi) \, \delta(E_{\rm obs} - g E_{\rm line})  \,  {\text{d}\alpha \, \text{d} \beta}.
\eeq 
% where we have included a general disc emissivity function $\epsilon(r, \phi)$ to allow for the possibility of an asymmetric illumination of the disc which varies with radius.
{Following e.g., \cite{Dauser2013}, we represent the emissivity as a broken power-law function of radius
\beq
\epsilon(r,\phi) \propto 
\begin{cases}
r^{-\gamma_{\rm in}}, \quad  r \leq r_{\rm br} \\
r^{-\gamma_{\rm out}}, \quad  r >    r_{\rm br},
\end{cases}
\eeq
where the inner and outer emissivity indices, $\gamma_{\rm in}$ and $\gamma_{\rm out}$, and the break radius, $r_{\rm br}$, are model parameters in \skl and \skc. This parameterisation is expected theoretically for a compact illuminating source \citep{Dauser2013,Ingram2019}. Whereas an extended source can instead require a twice broken power law \citep{Wilkins2012}. While twice broken power laws are indeed inferred observationally \citep{Wilkins2011}, a once broken power law provides a reasonable approximation and is widely used. The outer emissivity index is always expected to tend to $\gamma_{\rm out}=3$ for a flat accretion disc, whereas the inner index is $\gamma_{\rm in} \approx 2$ in Newtonian gravity \citep{Zycki1999} and $\gamma_{\rm in}>3$ in General Relativity \citep{Wilkins2012}. Values as high as $\gamma_{\rm in}\approx 5-10$ are routinely inferred observationally for the inner emissivity index \citep[e.g.,][]{Wilms2001,Miller2002,Wilkins2011,Dauser2012}. For the rest of this paper we simply employ $\gamma_{\rm in}=\gamma_{\rm out}\equiv\gamma$ (i.e., no break in the emissivity) to explore the line profile properties, and note that the effects driven by the absence of an event horizon will be stronger for a steeper emissivity profile.} We shall also assume that the disc extends down to $r_{\rm in} = r_I$ for the remainder of this paper.  Extensions of these models beyond the ISCO is naturally of interest, but the rapidly changing density of the intra-ISCO region \citep[e.g.,][]{MummeryBalbus2023} makes a more detailed treatment of the rest-frame spectrum important, and beyond the scope of the analysis of this paper. We will discuss the intra-ISCO region further in the conclusions of this paper.

\begin{figure}
    \centering
    \includegraphics[width=0.5\textwidth]{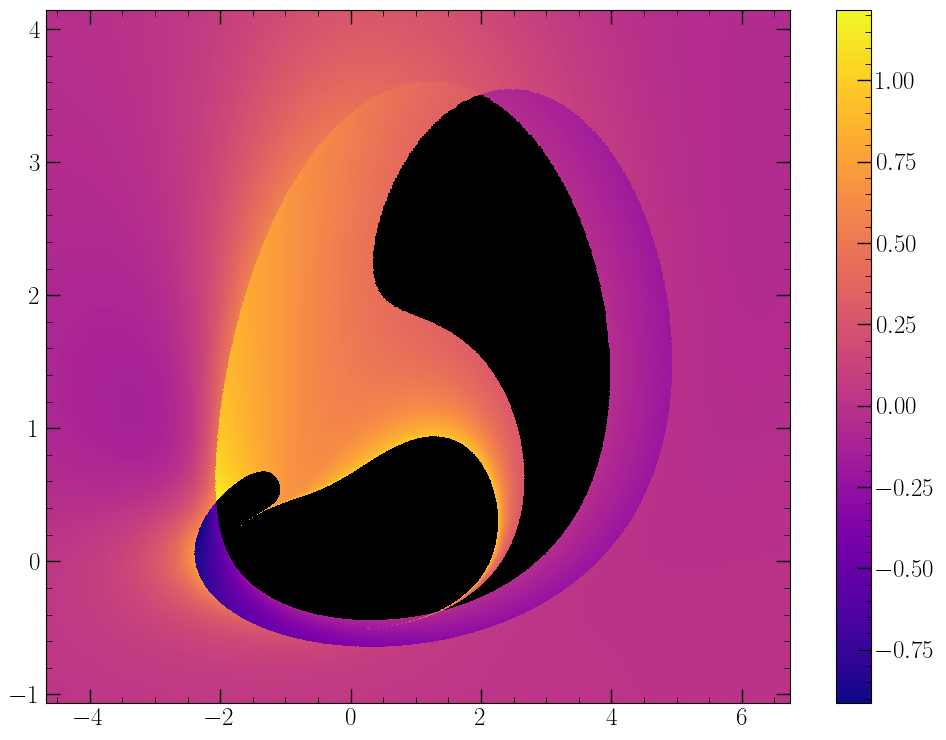}
    \caption{The difference in the energy shifts of photons observed in the image plane for a naked singularity system with $a_\bullet = 1.968$ and a black hole system with $a_\bullet = 0.998$. Both of these metrics have ISCO radii at the same location $r_I/r_g \simeq 1.237$, and both systems are observed at $\theta_{\rm obs} = 75^\circ$.  }
    \label{fig:delta-raytrace}
\end{figure}

\begin{figure}
    \centering
    \includegraphics[width=0.5\textwidth]{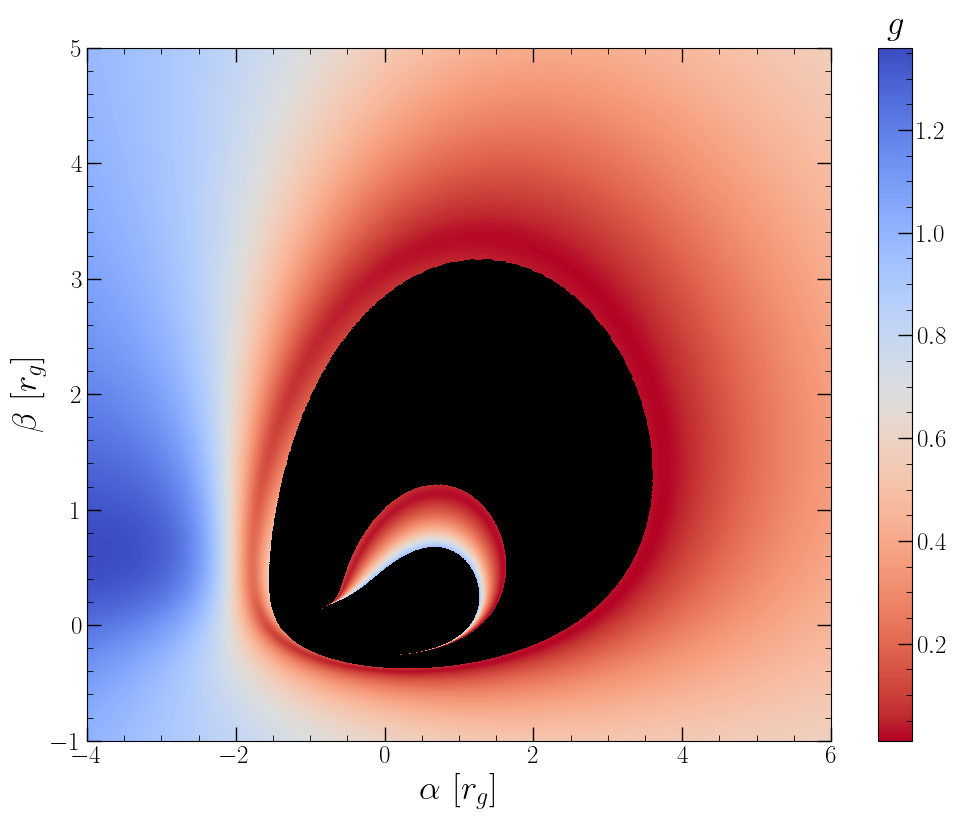}
    \includegraphics[width=0.5\textwidth]{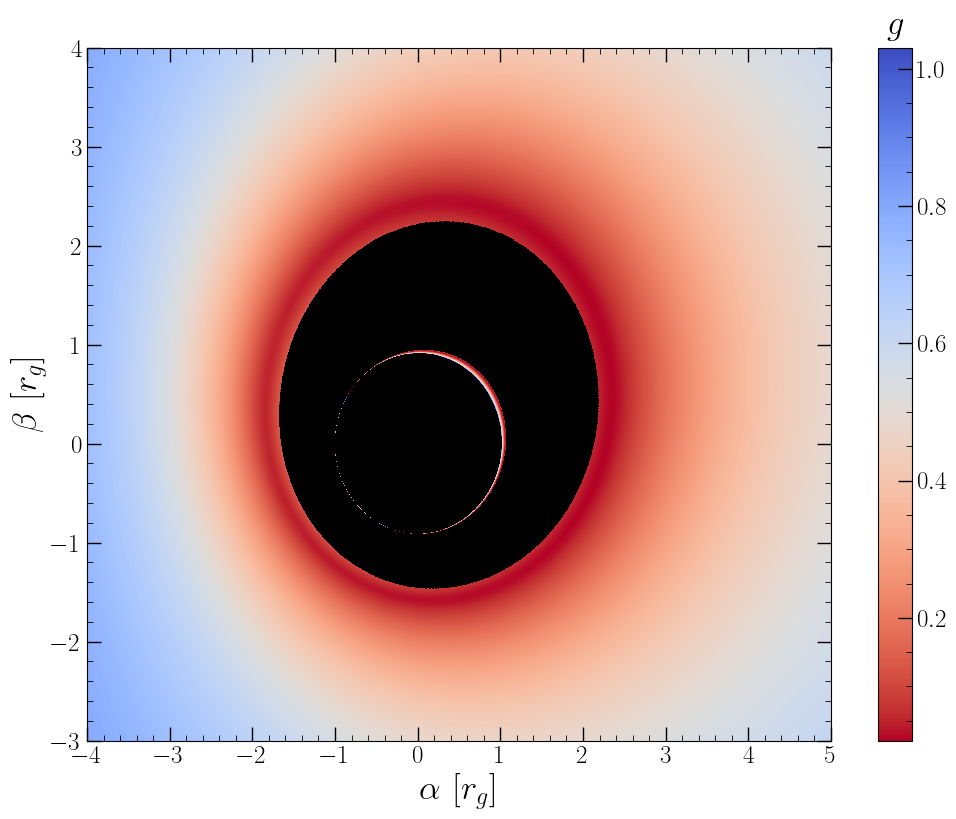}
    \caption{Image plane structure of the $a_\bullet = 1.002$ Kerr metric, observed at $\theta_{\rm obs} = 75^\circ$ (upper) and $\theta_{\rm obs} = 25^\circ$ (lower). }
    \label{fig:instant-structure-change}
\end{figure}

\section{ Photon energy shift maps  }\label{image-plane-maps}

The computation of $g(\alpha, \beta)$, which we shall henceforth refer to as determining the ``energy-shift map'',  must be performed numerically. One starts with a finely spaced grid $(\alpha, \beta)$ of photon impact parameters, at some large distance $d$ from the disc. This image plane is orientated at an angle $\theta_{\rm obs}$ from the singularity's spin-axis. For each photon,  the ray is traced back from the observer to the disc plane by solving the null-geodesics of the Kerr metric.  At the location at which the photon crosses the $\theta = \pi/2$ plane (i.e., we assume that the disc has infinitesimal thickness perpendicular to the spin axis), the radial coordinate $r$ and photon 4-momenta components $p_\phi, p_t$ are recorded. These three numbers fully specify the energy shift factor $g$.  Any photons lost into the event horizon (for black hole spacetimes), or which pass between the singularity and ISCO are not included. 

To perform these calculations we use  a modified version of the code \verb|YNOGK| \citep{YangWang13}, which is based on the code \verb|GEOKERR|: \citep{DexterAgol09}. We modify the \verb|YNOGK| code so as to compute null geodesics in the super-extremal Kerr spacetime \cite[see Appendix B of][]{MummeryBalbusIngram24}.

In Fig. \ref{fig:different-images} we display two photon energy shift maps, computed for a black hole spacetime with $a_\bullet = 0.998$ (upper row), and for a naked singularity metric with $a_\bullet = 1.968$ (lower row). This second spin value was chosen as it has an ISCO radius equal to that of the $a_\bullet=0.998$ black hole spacetime $r_I/r_g \simeq 1.237$. Both systems are observed at an inclination of $\theta_{\rm obs}= 75^\circ$. We include those photons which cross the disc plane with radii $r_I < r < 25r_g$.  The colour-bar in both cases displays the disc-observer energy shift of a photon observed at a given $\alpha, \beta$. Note that the normalisation of the two (upper and lower) colour-bars are identical.  

While on larger scales $|\alpha|, |\beta| \gtrsim 5r_g$ the energy shifts of the two metrics are comparable, there is a clear difference in the central regions, which we highlight explicitly in the right hand column. There are subtle differences in the ``normal'' region of the image plane, with the naked singularity metric having systematically bluer energy-shifts (because there is less gravitational red-shifting in the absence of an event horizon), but the main difference between the two is the obvious inner structure of the naked singularity image plane which is absent for black hole metrics.

This inner structure is entirely a result of the disappearance of the event horizon for Kerr metrics satisfying $|a_\bullet| > 1$.  The structure occurs due to photons from the disc passing within $r \sim 1r_g$ of the singularity and, instead of being swallowed by an event horizon, are ``spat out'' of the region and redirected towards the observer. The photons in this region of the image were never in the $\theta < \pi/2$ region (no such photons contribute to the image, since we assume they will be absorbed by optically thick material), they were emitted from the upper surface of the disc and ``bounced'' off the naked singularity. We will show that this image plane structure has a profound impact on observed iron line structures of naked singularity metrics. 

The exact difference in the energy shift maps, defined as 
\begin{equation}
    \Delta g(\alpha, \beta) \equiv g_{\rm naked \, singularity} - g_{\rm black \, hole},  
\end{equation}
is displayed in Fig. \ref{fig:delta-raytrace}. Despite sharing a common ISCO location, these two metrics are very different in the innermost regions of the image plane.

This additional image plane structure is not limited to super-extremal Kerr metrics with $a_\bullet \gg 1$, it in fact occurs for any Kerr metric with $a_\bullet > 1$ and for all observing inclinations. To verify this see Fig. \ref{fig:instant-structure-change}, where we display the image plane structure of the $a_\bullet = 1.002$ Kerr metric, observed at $\theta_{\rm obs} = 75^\circ$ (upper panel) and $\theta_{\rm obs} = 25^\circ$ (lower panel). The upper panel shows that for larger inclinations the change in image plane structure is pronounced across the extremal $a_\bullet = 1$ limit.  This change is much less pronounced for near face-on inclinations (lower panel), but nonetheless a new inner ring of observable photons still exists. In the following section we examine the effects of these results on the observed line profiles originating from Kerr metric discs. 

\section{ Observed line profiles }\label{line-profiles}

The differing image plane structures of black hole and naked singularity Kerr metrics result in extreme differences in the properties of observed line profiles, for much of parameter space. {Here we explore these differences under the assumption that the disc inner radius is at the ISCO.}  

\begin{figure}
    \centering
    \includegraphics[width=.5\textwidth]{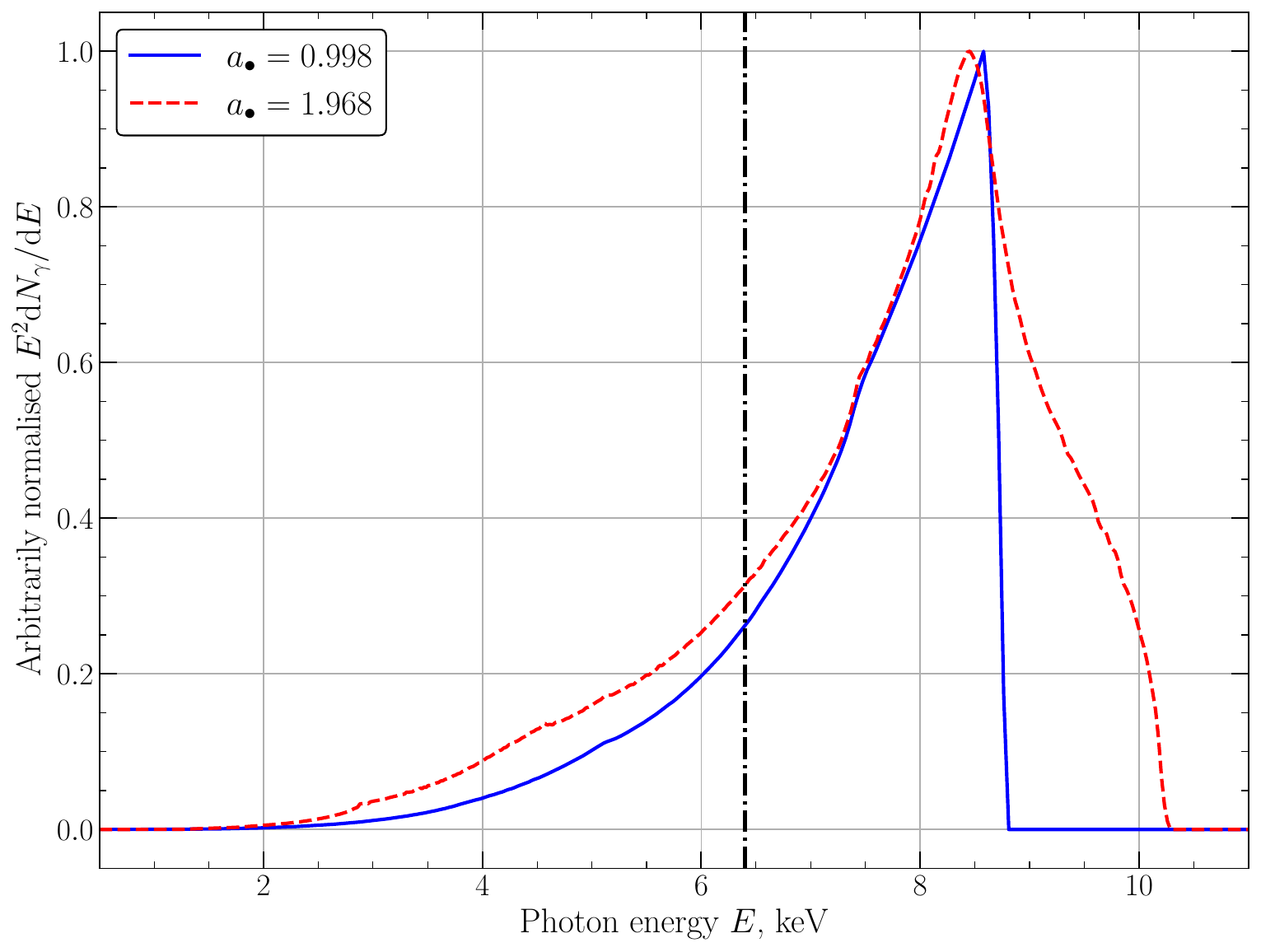}
    \caption{The very different observed line profiles of two Kerr metric discs which share a common ISCO location. The solid blue line profile is produced for a black hole metric with $a_\bullet = 0.998$, with ISCO $r_I/r_g \simeq 1.237$, while the red dashed line profile is for a naked singularity metric with $a_\bullet = 1.968$, and identical ISCO $r_I/r_g \simeq 1.237$ location. The other parameters of the system are: $\theta_{\rm obs} = 75^\circ$, $\gamma = 3.0$, $E_{\rm line} = 6.4$ keV (black dot-dashed line) and $r_{\rm out} = 30r_g$. The differing properties of the observed line profiles can be understood with reference to the observer image plane properties of the two metrics (Fig. \ref{fig:different-images}; see text). }
    \label{fig:different-lines}
\end{figure}

As a first example of this, see Fig. \ref{fig:different-lines}. In this Figure we plot line profiles of two discs which share a common ISCO location  $r_I/r_g \simeq 1.237$, one for a black hole metric $a_\bullet = 0.998$ (blue solid curve) and another for a naked singularity metric $a_\bullet = 1.968$ (red dashed curve). The line profile is plotted as $E^2 {\rm d}N_\gamma / {\rm d}E$, where $E$ is the photon energy and $N_\gamma$ is the number of detected photons, the normalisation of these profiles is arbitrary. Both systems were observed at an inclination $\theta_{\rm obs} = 75^\circ$ (i.e., they had photon energy shift maps of Fig. \ref{fig:different-images}), and an emissivity index $\gamma = 3$. The rest-frame line energy was taken to be that of the Fe K$\alpha$ transition $E_{\rm line} = 6.4$ keV, and we include emission from the inner 30$r_g$ of the disc. The naked singularity metric results in a significantly broader observed line, with red and blue wings spreading to lower and higher energies than the black hole metric. 

The extension to higher energies (the blue wing) results directly from the lack of an event horizon.  Generically it is  gravitational red-shifting which works against the Doppler boosting of the disc fluids rotation and ultimately limits the extent of the blue wing. The lack of event horizon reduces the effects of gravitational red-shifting, and allows for an extended blue wing despite the rotational disc velocities of the two metrics being similar.  

The observed extension of the red wing results from the larger observed area of the Doppler red-shifted disc regions (Fig. \ref{fig:different-images}). Again, this is a direct result of the lack of an event horizon. 

\begin{figure}
    \centering
    \includegraphics[width=.5\textwidth]{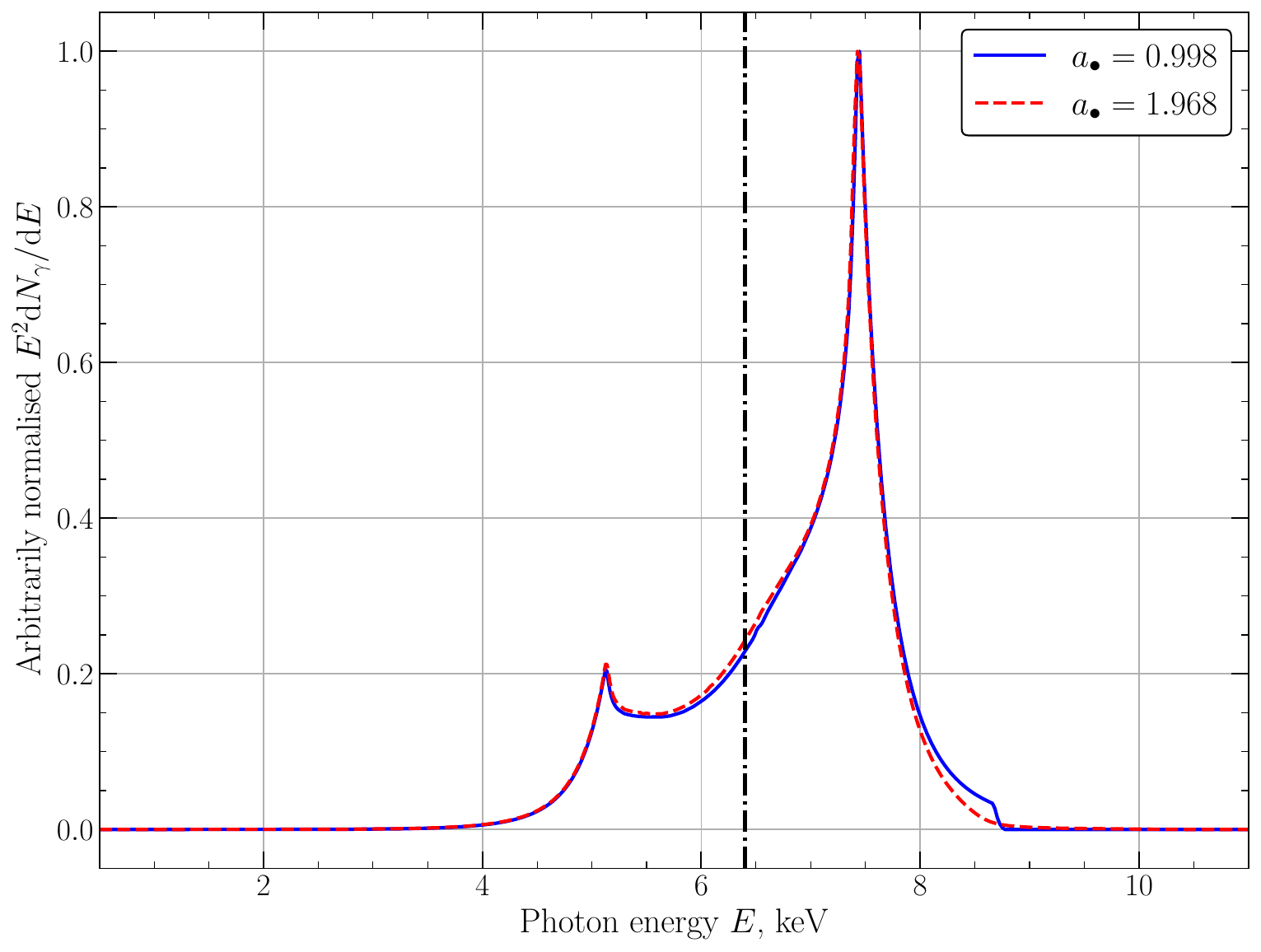}
    \caption{Similar observed line profiles of two Kerr metric discs which share a common ISCO location. All parameters are identical to Figure \ref{fig:different-lines}, except the emissivity index which now equals $\gamma = 0.0$. The shallower illumination profile ($\gamma=0$) reduces the importance of the event horizon, resulting in similar results for the two metrics. Emissivity profiles this shallow are likely unphysical, and inferred indices from spectral fitting are always much steeper. }
    \label{fig:similar-lines}
\end{figure}

\begin{figure}
    \centering
    \includegraphics[width=.5\textwidth]{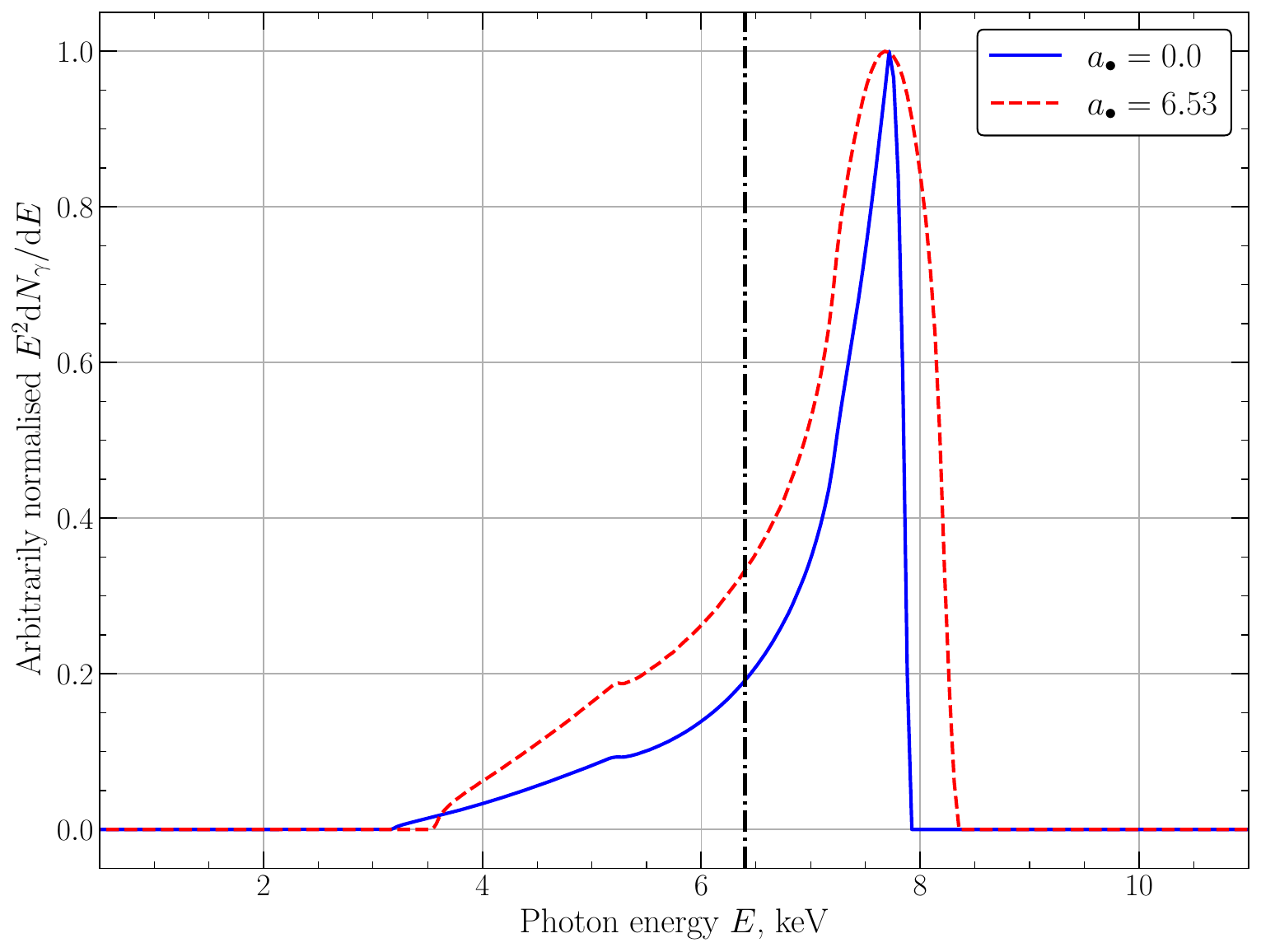}
    \caption{ The observed line profiles of the Schwarzschild metric, and a naked singularity metric (with $a_\bullet = 6.53$) which shares the same ISCO location.  The other parameters of the system are: $\theta_{\rm obs} = 60^\circ$, $\gamma = 4.0$, $E_{\rm line} = 6.4$ keV (black dot-dashed line). The two metrics produce noticeably different line profiles.  }
    \label{fig:schwarz-like-lines}
\end{figure}

\begin{figure}
    \centering
    \includegraphics[width=.5\textwidth]{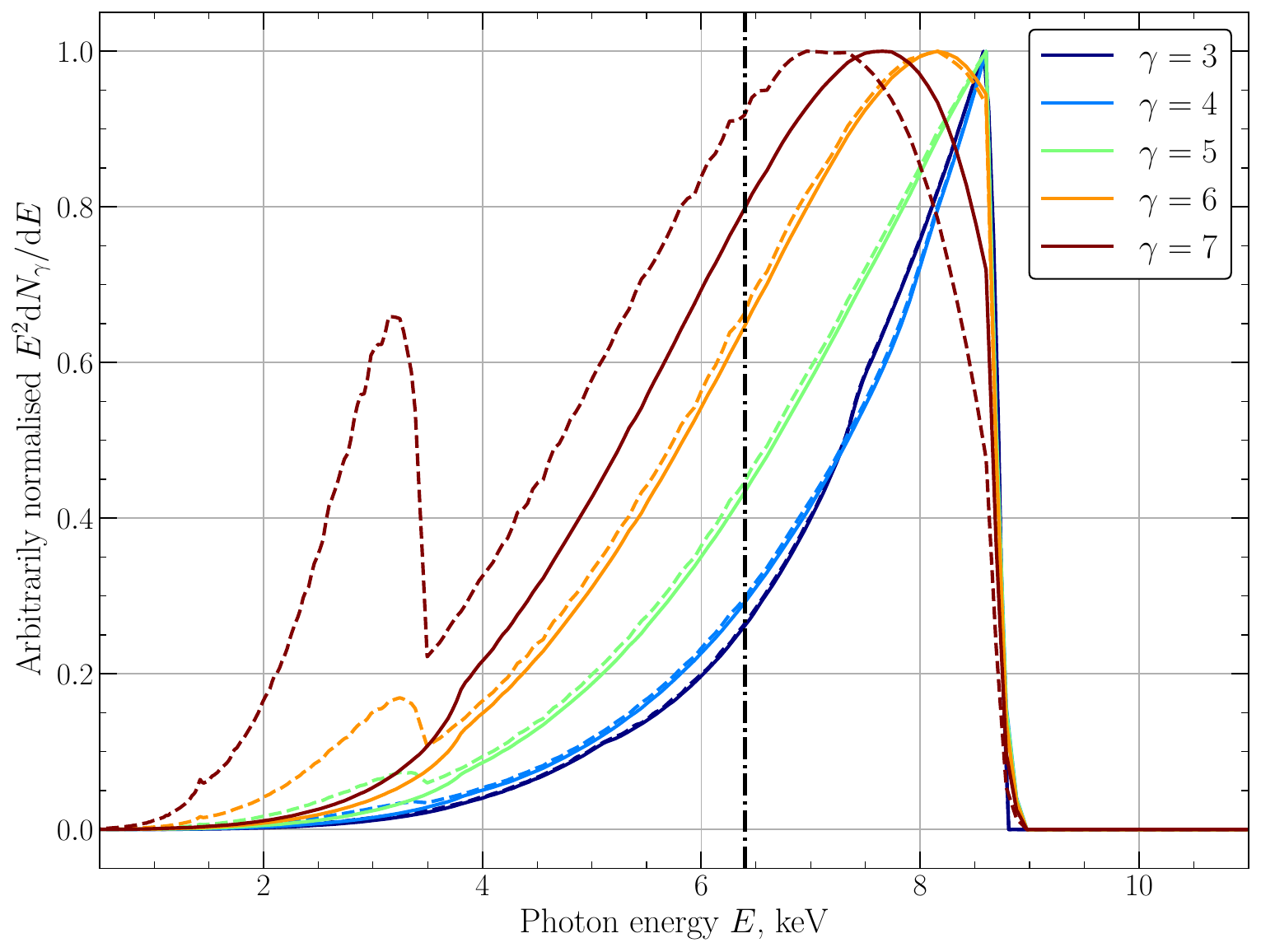}
    \caption{ Observed line profiles from two Kerr metrics, one describing a rapidly rotating black hole $a_\bullet = 0.998$ (solid curves), and another for a just-super extremal $a_\bullet = 1.002$ metric (dashed curves), for differing emissivity indices $\gamma$ (denoted by different line colours). For shallow emissivity profiles $\gamma \lesssim 5$ the line profiles are similar, but for steep emissivity profiles they are clearly distinguishable due to the emergence of double-lines in the naked singularity spectrum. For these line profiles $\theta_{\rm obs} = 75^\circ$.    }
    \label{fig:gamma_dep_lines}
\end{figure}

It is only for relatively steep (i.e., $\gamma \gtrsim 2$) emissivity profiles that these differences become apparent. This can be seen with reference to Fig. \ref{fig:similar-lines}, which shows lines produced with identical system parameters to Fig. \ref{fig:different-lines} except the emissivity index $\gamma$, which is now set to $\gamma = 0$ (i.e., $\epsilon(r) = {\rm cst}.$). The line profiles in this case are dominated by Doppler shifting of photons produced at larger radii (with larger emitting areas). At these large radii the two metrics are near identical in the image plane (see Fig. \ref{fig:different-lines}), and the importance of the event horizon is minimal. Fortunately for the purposes of probing the cosmic censorship conjecture, emissivity profiles this shallow are likely unphysical, and inferred indices from spectral fitting are typically much steeper \citep[e.g.][]{Wilms2001,Miller2002,Wilkins2011,Dauser2012}.

It is not just for black hole metrics with large spins that the differences from the naked singularity metrics with shared ISCO locations become apparent, the same is true for all black hole metrics.  This can be seen in Fig. \ref{fig:schwarz-like-lines}, where we plot the observed line profiles of a Schwarzschild $a_\bullet =0$ metric, and a naked singularity metric with $a_\bullet = 8\sqrt{6}/3 \approx 6.53$, which has an identical ISCO location at $r_I = 6r_g$. The other parameters of the system are: $\theta_{\rm obs} = 60^\circ$, $\gamma = 4.0$, $E_{\rm line} = 6.4$ keV (black dot-dashed line). The two metrics produce noticeably different line profiles.

Even Kerr metrics described by $a_\bullet = 1 \pm \delta, (\delta \ll 1)$ can be distinguished for sufficiently steep emissivity indices $\gamma$. In Fig. \ref{fig:gamma_dep_lines} we plot the observed line profiles from two Kerr metrics, one describing a rapidly rotating black hole $a_\bullet = 0.998$ (solid curves), and another for a just-super extremal $a_\bullet = 1.002$ metric (dashed curves), for differing emissivity indices $\gamma$ (denoted by different line colours). For shallow emissivity profiles $\gamma \lesssim 5$ the line profiles are similar, but for steep emissivity profiles they are clearly distinguishable due to the emergence of double-lines in the naked singularity spectrum.  

It is noteworthy that the continuum emission of thin discs evolving in metrics with $a_\bullet = 1\pm \delta$ are radically different \citep[as shown in][]{MummeryBalbusIngram24}, and so X-ray binaries with both observed line profiles and continuum emission  {\it should} easily be able to distinguish between black hole and naked singularity metrics.  

\begin{figure}
    \centering
    \includegraphics[width=.47\textwidth]{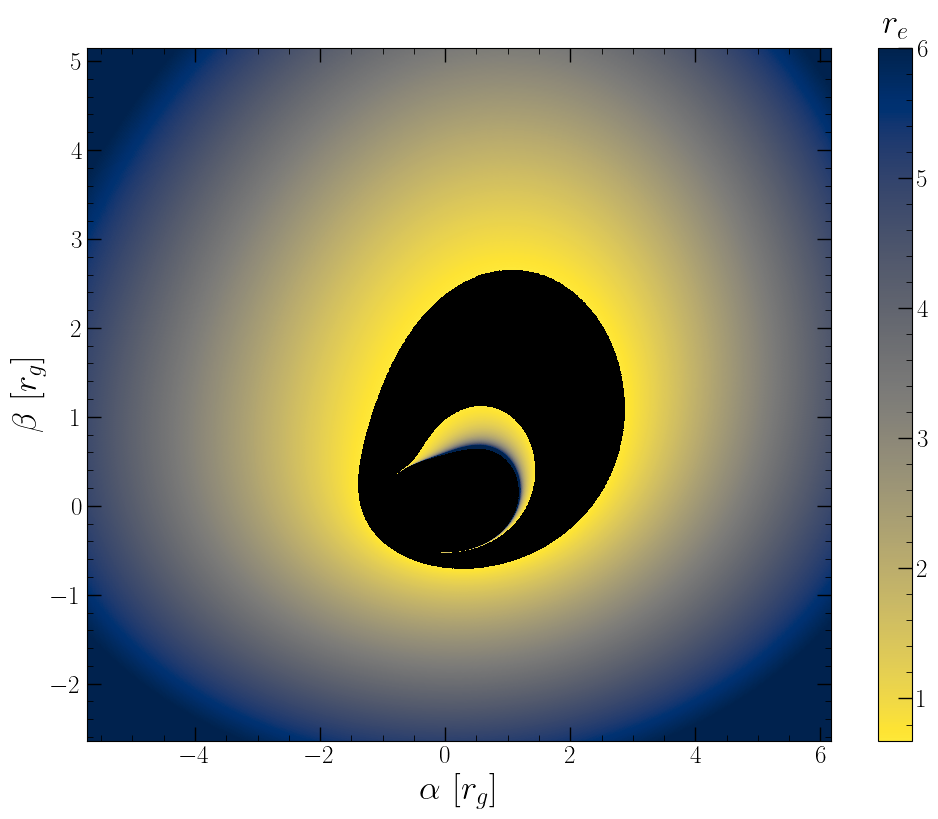}
    \includegraphics[width=.47\textwidth]{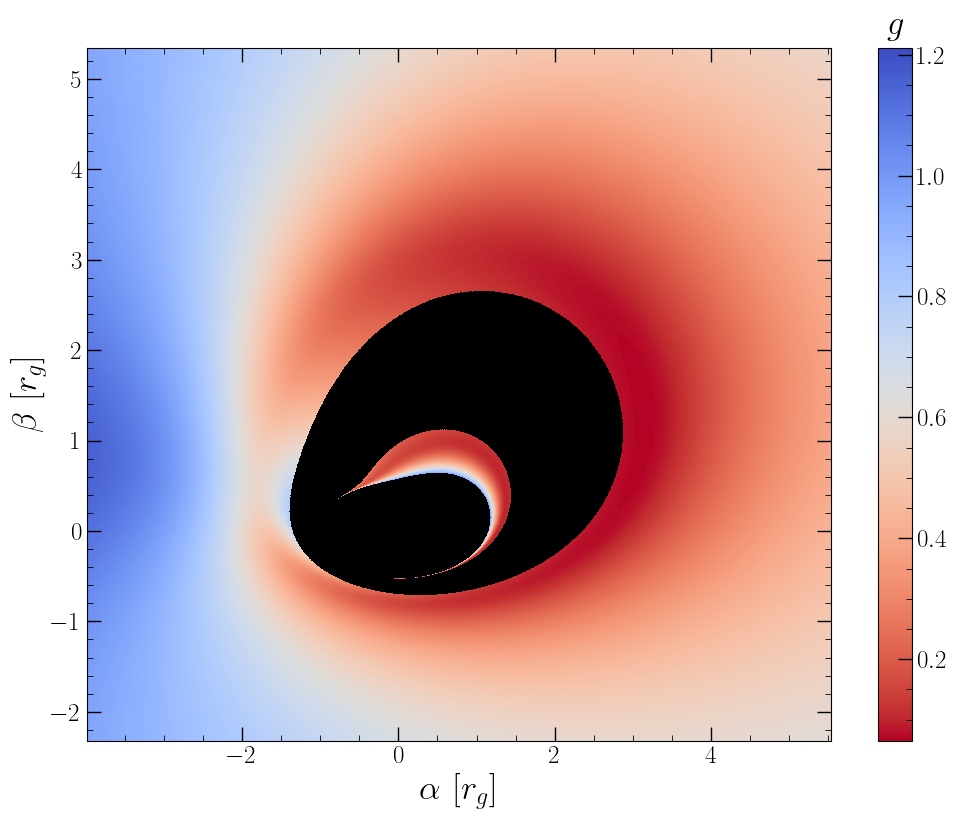}
    \includegraphics[width=.45\textwidth]{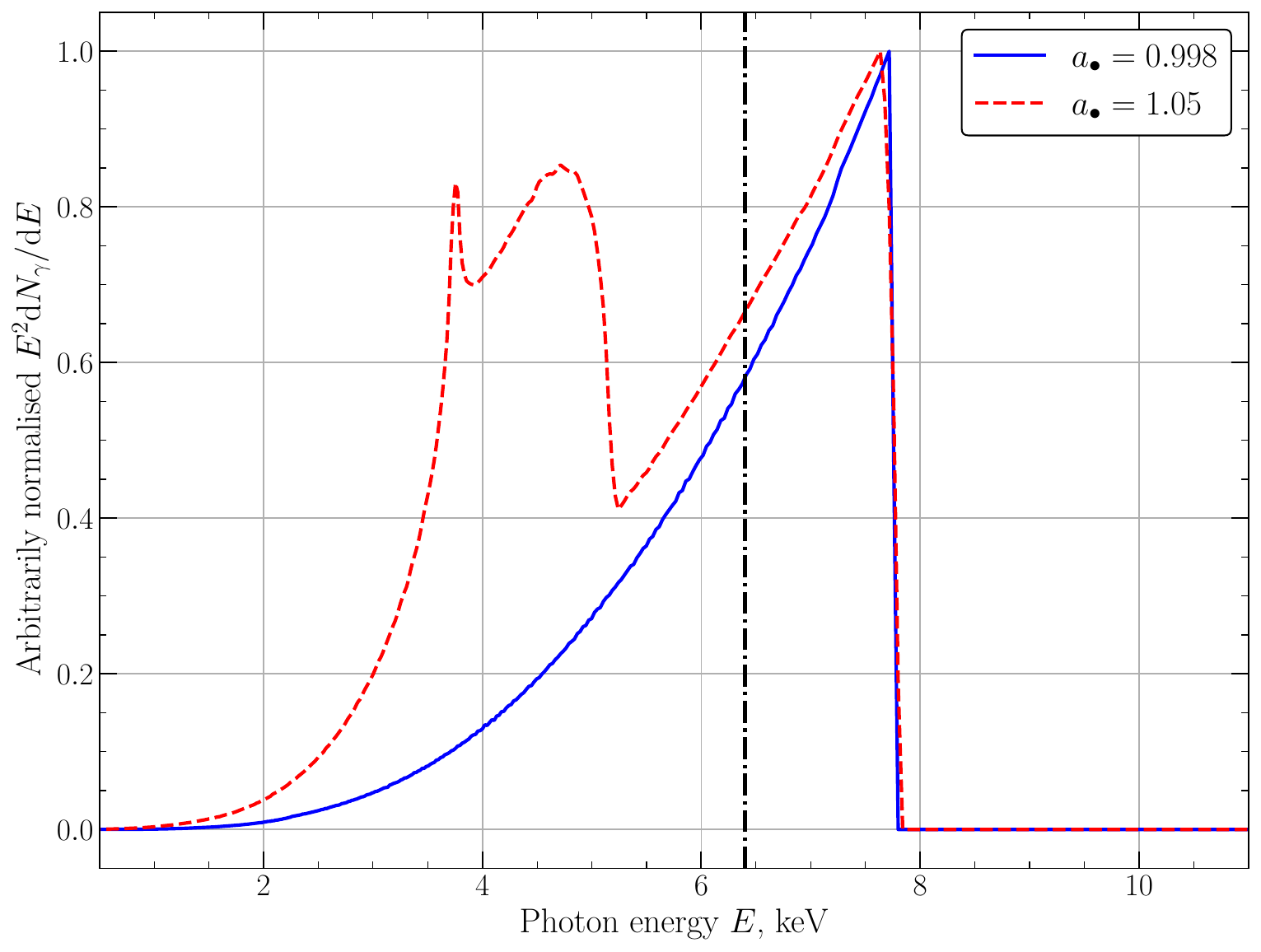}
    \caption{ Lowest panel, red curve: an example of a ``triple''  naked singularity iron line. Physical parameters $a_\bullet = 1.05, \theta_{\rm obs} = 60^\circ, \gamma = 4$. The central cluster of photons in the image plane originate from the inner edges of the disc (photon emission radius displayed in top panel), and contributes in effect a second red-shifted line (energy shifts displayed in middle panel), in addition to the ``normal'' disc line. This second line is then itself split by Doppler boosting of the inner photon cluster, resulting in a triple line.  For reference in the lowest panel we display in blue an $a_\bullet = 0.998$ line, with other parameters identical.  }
    \label{fig:mad-lines}
\end{figure}

The emergence of ``double-lines'' in naked singularity spectra are generic properties of Kerr metrics with $a_\bullet = 1 + \delta$ ($\delta \ll 1$)\footnote{Of course, black hole metrics with emission dominated from large radii also show double lines, driven entirely from Doppler shifting (e.g., Fig. \ref{fig:similar-lines}). These double lines show some symmetry about $E_{\rm line}$ however, which is very different from the double gravitationally shifted lines we are describing here.  }. In fact, it is even possible to generate ``triple'' observed lines (from a single delta-function rest-frame line), as demonstrated in Fig. \ref{fig:mad-lines}.  The lowest panel of this Figure shows an observed emission line from a system with physical parameters $a_\bullet = 1.05, \theta_{\rm obs} = 60^\circ, \gamma = 4$. The inner cluster of photons (see upper two panels) originate from the inner edges of the disc (photon emission radius is displayed in the top panel), and contribute in effect a second red-shifted line (energy shifts displayed in middle panel), in addition to the ``normal'' disc line. This second line is then itself split by Doppler boosting of the disc regions which produce the inner photon cluster, resulting in a triple line.   For reference in the lowest panel we display in blue an $a_\bullet = 0.998$ line, with other parameters identical.  The fact that the inner cluster of photons originate from the very innermost disc regions (upper panel Fig. \ref{fig:mad-lines}) highlights why these doublet (and triplet) lines appear for steep emissivity profiles $\gamma$, when the lines are entirely dominated by emission from the inner $\sim r_g$ of the disc. 

\begin{figure*}
    \centering
    \includegraphics[width=0.49\textwidth]{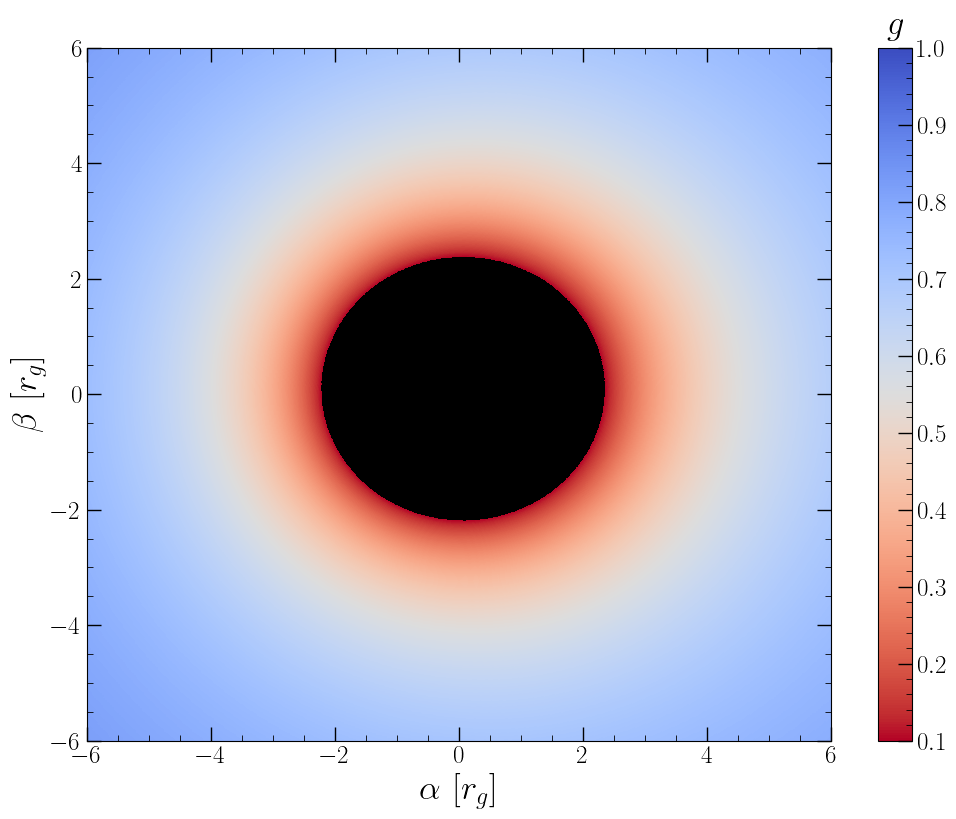}
    \includegraphics[width=0.49\textwidth]{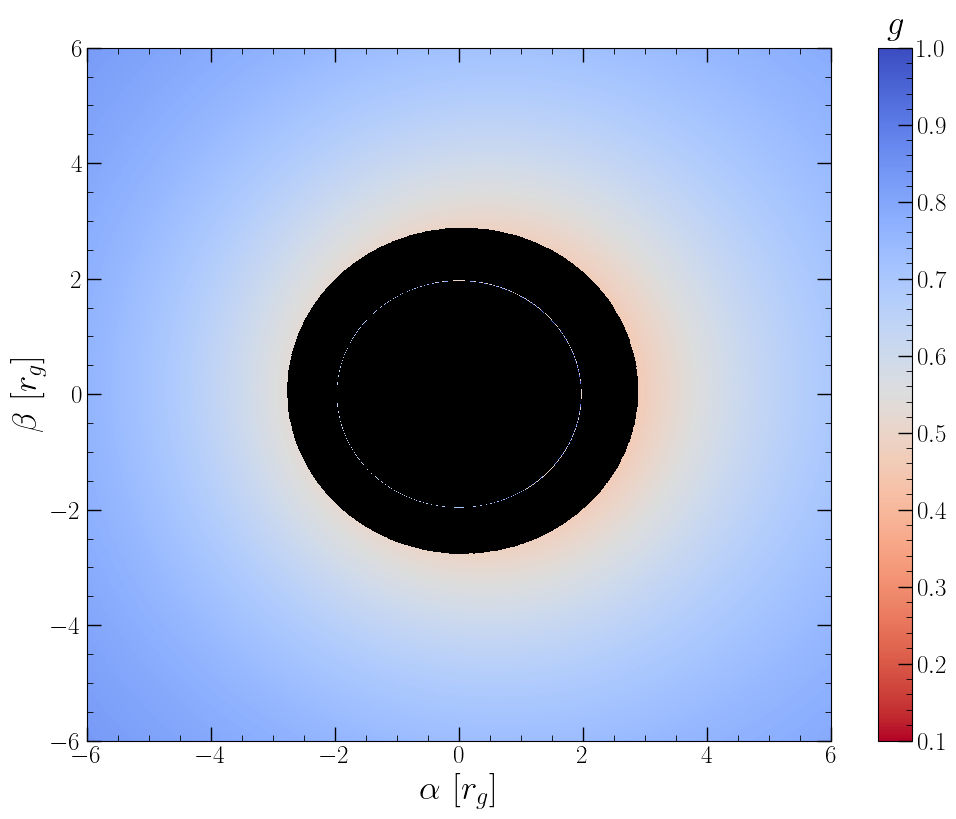}
    \includegraphics[width=0.49\textwidth]{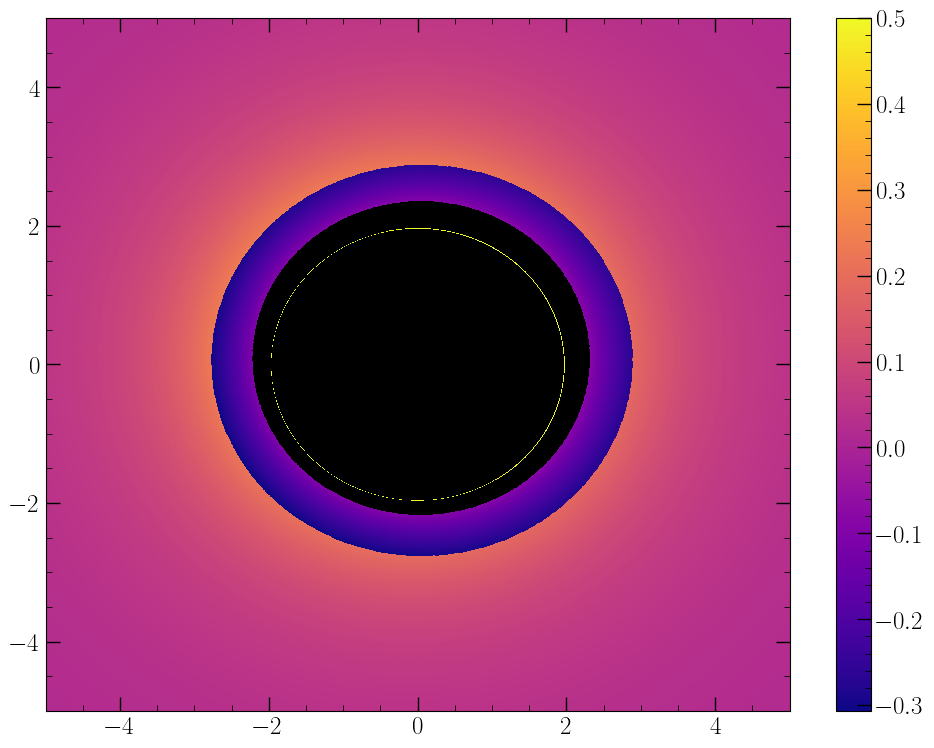}
    \includegraphics[width=0.49\textwidth]{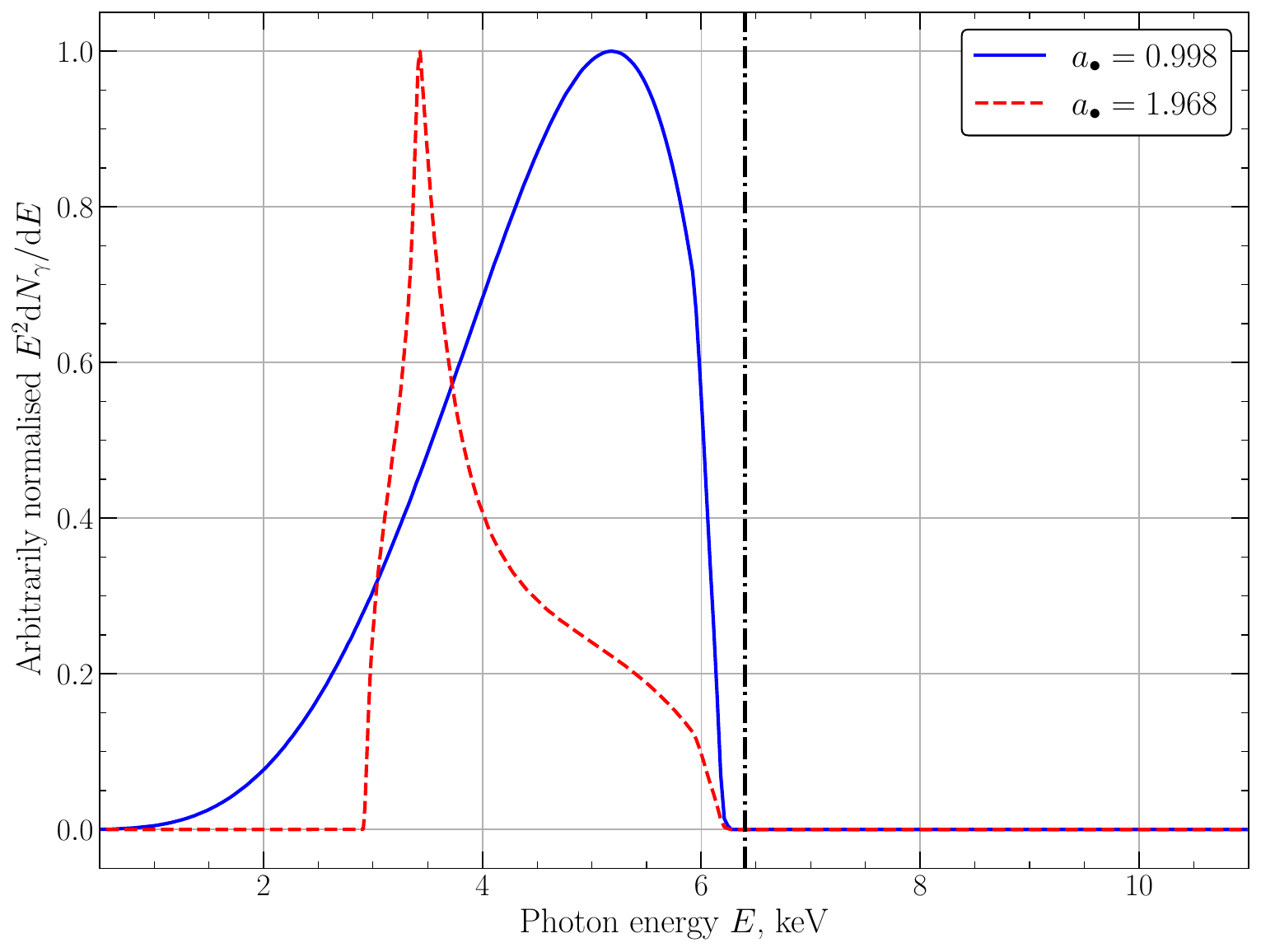}
    \caption{ Lower right panel: inverted iron lines from a near face-on observation of a naked singularity disc. In red (dashed) we display the observed emission line from a $a_\bullet = 1.986$ metric, and in blue (solid) we show a $a_\bullet = 0.998$ metric. Other physical parameters $\theta_{\rm obs} = 5^\circ, \gamma = 4$. The much reduced gravitational red-shifting of the naked singularity metric results in this inverted profile. Upper left panel: energy-shift map of $a_\bullet = 0.998$ metric. Upper right panel: energy-shift map of $a_\bullet = 1.968$ metric (both energy shift colour bars are normalised to show the same scale). Lower left panel: the difference in energy shift maps, defined by $\Delta g(\alpha, \beta) \equiv g_{\rm naked \, singularity} - g_{\rm black \, hole}$. }
    \label{fig:face-on-inverted-line}
\end{figure*}

Another bizarre feature of naked singularity X-ray spectral lines is highlighted in Fig. \ref{fig:face-on-inverted-line}. For face-on discs evolving in naked singularity Kerr metrics one can observe ``inverted'' iron lines.  We use the nomenclature ``inverted'' as a typical black hole metric iron line has a sharp blue edge with an extended red tail (e.g., Figs. \ref{fig:different-lines}, \ref{fig:schwarz-like-lines}, \ref{fig:gamma_dep_lines}, \ref{fig:mad-lines}). However, if a naked singularity disc is observed near face on, the observed line has a sharp red edge and an extended blue tail
% an observed line with sharp red-edge, and extended blue tail can be observed
(lower right panel Fig. \ref{fig:face-on-inverted-line}).  Once again, this unique feature is a result of the effect of the disappearance of the event horizon on the energy shift maps of the naked singularity metric.  In the upper left panel we display the energy shift map of the $a_\bullet = 0.998$ black hole metric, observed at $\theta_{\rm obs} = 5^\circ$, while in the upper right panel we display the energy shift map of the $a_\bullet = 1.968$ naked singularity metric, also observed at $\theta_{\rm obs} = 5^\circ$. While the two energy shift maps look  similar, the lack of an event horizon means that the gravitational energy shifts of the $a_\bullet = 1.968$ metric are substantially less than the $a_\bullet = 0.998$ metric, despite them sharing a common ISCO location. The difference in the two energy shifts is displayed in the lower left panel of Fig. \ref{fig:face-on-inverted-line}, where we see a systematic energy shift  difference for the inner disc radii of order $\Delta g \approx 0.3$. This, coupled with the larger observed area of the inner disc in the $a_\bullet = 1.968$ metric, results in the spectral inversion seen in the lower right panel.  

\section{ Convolved X-ray spectra }\label{conv}
\begin{figure}
    \centering
    \includegraphics[width=\linewidth]{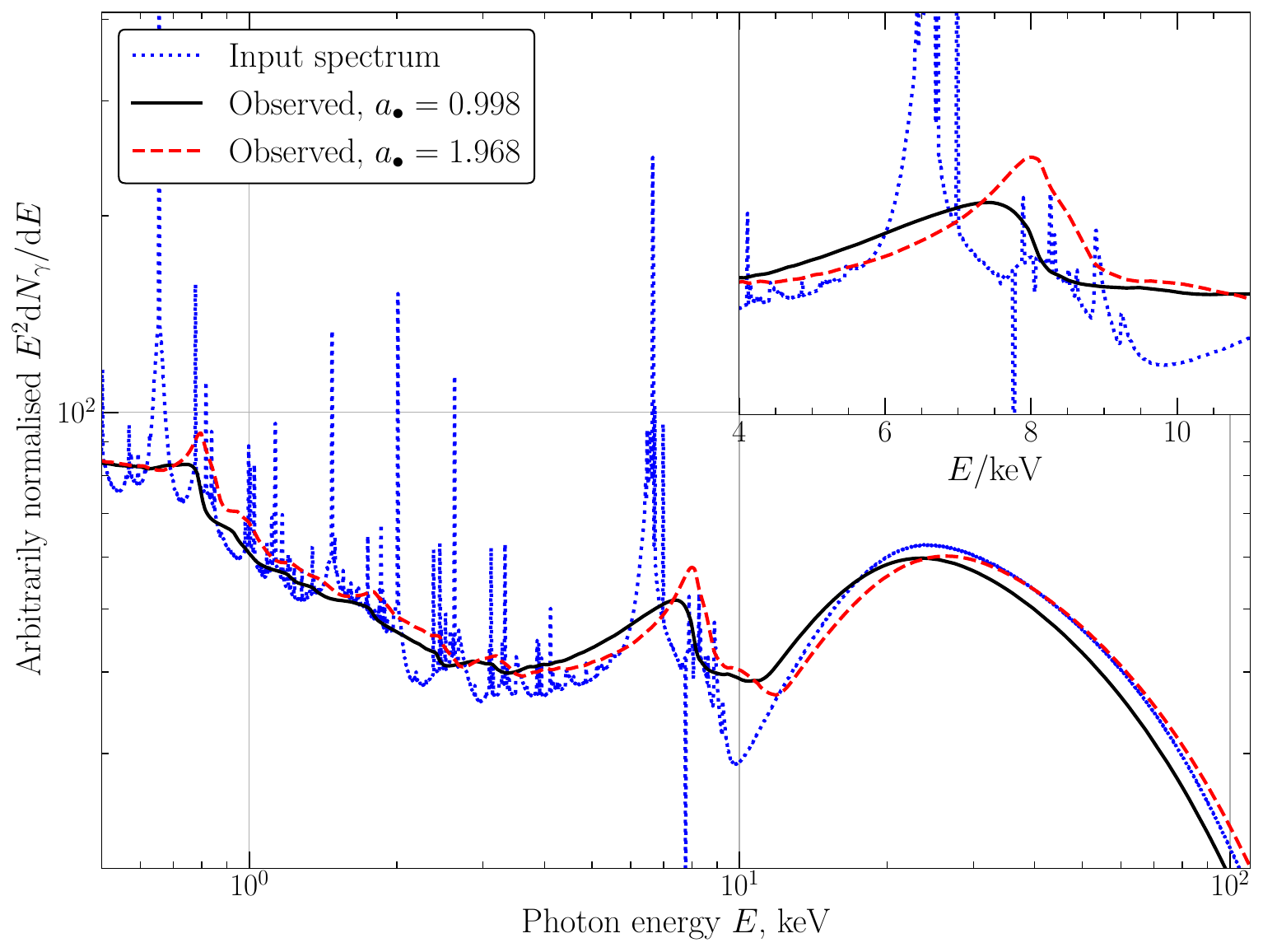}
    \caption{Two convolved X-ray spectra produced from observing a single rest frame spectrum in two Kerr metrics with differing spin parameters.  The rest frame spectrum (blue dotted line) is calculated with \texttt{xillver} with default parameters ($\Gamma=2$, $\log_{10}\xi=3.1$, $A_{\rm Fe}=1$, $E_{\rm cut}=300$ keV). We convolve with \skc for two spin values, $a_\bullet=0.998$ (black solid line) and $a_\bullet=1.968$ (red dashed line). The other parameters are $\theta_{\rm obs} = 75^\circ$, $\gamma = 4$. We see that these two examples, which have the same ISCO radius, are still very different from one another. The inset in the upper right shows a zoom-in around the FE K$\alpha$ line, showing that these two metrics are distinguishable.  }
    \label{fig:convolved}
\end{figure}

Consider a general rest-frame X-ray spectrum $I_E(E)$ emitted from a disc location $(r, \phi)$. We will assume for simplicity that this rest frame spectrum factorises into a disc-location independent shape function $I_E(E)$ and a disc location dependent amplitude (or emissivity) $\epsilon(r, \phi)$. Under this assumption, the emitted rest frame spectrum can always be written as an integral over a delta function 
\begin{equation}
    I_E(E,r , \phi) \propto \epsilon(r, \phi) \int_0^\infty I_E(E') \, \delta(E-E') \, {\rm d}E' ,
\end{equation}
and, following the procedure spelt out in section \ref{line-fitting-formalism}, the  flux observed by a distant observer at rest  will be 
% \beq
% F_{E}(E_{\rm obs} ) = \iint g^3 I_E(E_{\rm emit}) \, \text{d}\Omega,
% \eeq 
% or explicitly 
\beq
F_{E}\propto \iint g^3 \left[ \epsilon(r, \phi) \int_0^\infty I_E(E') \, \delta(E_{\rm obs}/g - E') \, {\rm d}E' \right]  \text{d}\Omega  .
\eeq 
Swapping the order of integration, we have 
\beq
F_{E}\propto \int_0^\infty I_E(E') \left[ \iint g^3 \epsilon(r, \phi)  \, \delta(E_{\rm obs}/g - E')  \, \text{d}\Omega \right]  {\rm d}E' .
\eeq 
{This can be represented as a convolution between the line profile (like that calculated above) and the rest-frame reflection spectrum \citep[see e.g.,][for further details]{Ingram2019}. Such a ``Green's function'' approach
is employed by the \skc model. Fig \ref{fig:convolved} shows an example. The rest frame spectrum (blue dotted line) is calculated with \texttt{xillver} \citep{Garcia2013} with default parameters ($\Gamma=2$, $\log_{10}\xi=3.1$, $A_{\rm Fe}=1$, $E_{\rm cut}=300$ keV). We convolve with \skc for two spin values, $a_\bullet=0.998$ (black solid line) and $a_\bullet=1.968$ (red dashed line).  The other parameters are $\theta_{\rm obs} = 75^\circ$, $\gamma = 4$. We see that these two examples, which have the same ISCO radius, are still very different from one another. Capturing the additional physics of (for example) a radial dependence of the rest-frame reflection spectrum requires extra computational expense and will be considered in the future. For now, we note that it is an important result that more complex convolved X-ray spectra can be distinguished even for discs with the same ISCO location.  }

% This is precisely the functional form of a convolution between the input spectrum $I_E(E')$ and the delta function line profiles computed in this paper. 

\begin{figure}
    \centering
    \includegraphics[width=\linewidth]{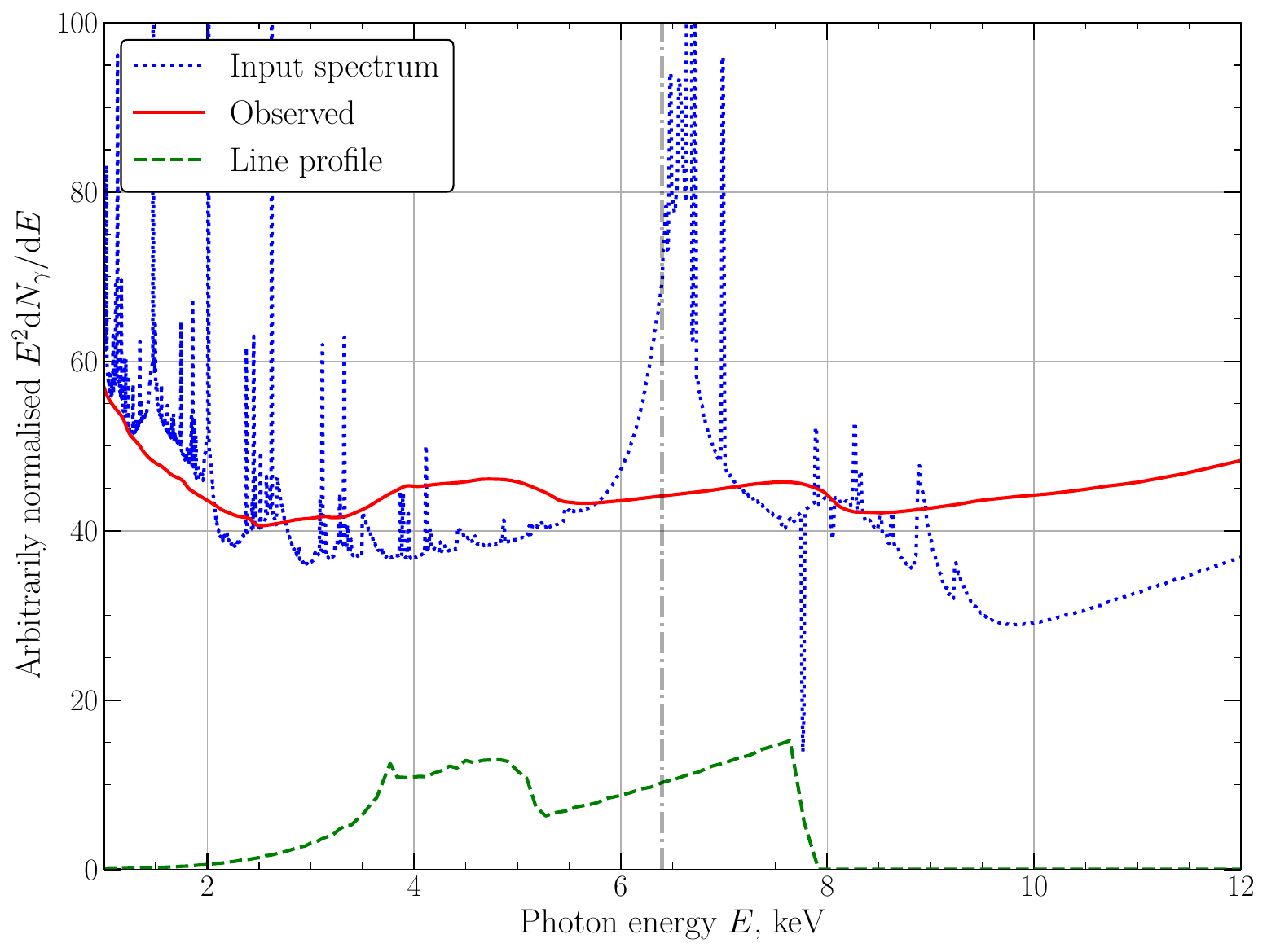}
    \includegraphics[width=\linewidth]{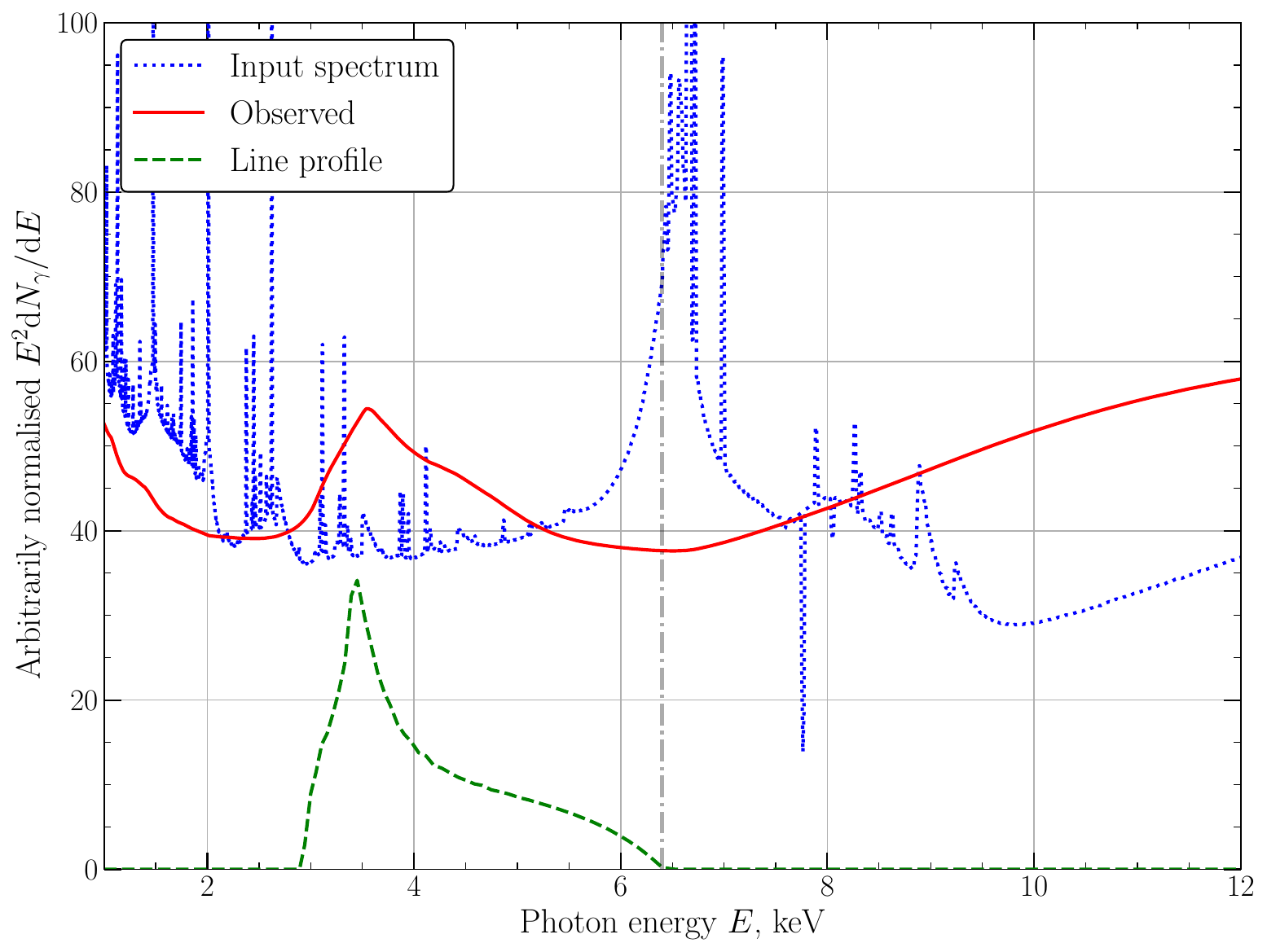}
    \caption{Two convolved X-ray spectra produced from observing a single rest frame spectrum in Kerr metrics with differing spin parameters.  The rest frame spectrum (blue dotted line) is calculated with \texttt{xillver} with default parameters ($\Gamma=2$, $\log_{10}\xi=3.1$, $A_{\rm Fe}=1$, $E_{\rm cut}=300$ keV). We convolve with \skc for two sets of parameters which produce notable line profiles in the super-extremal Kerr metric $a_\bullet= 1.05$, $\theta = 60^\circ$ (upper panel, see also Fig. \ref{fig:mad-lines}) and $a_\bullet=1.968$, $\theta_{\rm obs} = 5^\circ$ (lower panel, Fig. \ref{fig:face-on-inverted-line}). The emissivity index is $\gamma = 4$. Both of the extreme line profiles translate into clearly observable features in a more detailed reflection spectrum. The green dashed curves show the $\delta$-function line profiles, produced assuming a line energy $E_{\rm line} = 6.4$ keV. The normalisation for this line is arbitrary and is shown only for pedagogical purposes.     }
    \label{fig:convolved_interesting}
\end{figure}

In Figure \ref{fig:convolved_interesting} we display convolved X-ray spectra for two sets of parameters which produce notable line profiles in the super-extremal Kerr metric $a_\bullet= 1.05$, $\theta = 60^\circ$ (upper panel, see also Fig. \ref{fig:mad-lines}) and $a_\bullet=1.968$, $\theta_{\rm obs} = 5^\circ$ (lower panel, see also Fig. \ref{fig:face-on-inverted-line}). The rest frame spectrum (blue dotted curve) is identical to Fig. \ref{fig:convolved}. The emissivity index is $\gamma = 4$. Both of the extreme line profiles translate into clearly observable features in a more detailed reflection spectrum. The green dashed curves show the $\delta$-function line profiles, produced assuming a line energy $E_{\rm line} = 6.4$ keV. The normalisation for this line is arbitrary and is shown only for pedagogical purposes.  

\section{Discussion and conclusions}\label{conclusions}
In this paper we have highlighted how
% the fitting of
iron lines observed in the spectra of accreting X-ray binaries and Active Galactic Nuclei can be leveraged as powerful probes of the (weak) cosmic censorship conjecture \citep{Penrose1969}. While disc-continuum fitting (the other chief approach to inferring Kerr metric spin parameters) suffers from degeneracy between sub- and super-extremal metrics for some regions of parameter space \citep[e.g.,][]{Takashi10, MummeryBalbusIngram24}, the same is not true for iron line fitting.  The reason for this difference is that iron line fitting typically probes much smaller physical scales than continuum fitting, where the differences between black hole and naked singularity metrics are most pronounced.  

The small-radii image plane differences between black hole and naked singularity metrics can be traced directly to the disappearance of the Kerr metric event horizon for $|a_\bullet | > 1$. There are subtle differences in the ``normal'' region of the image plane, with  naked singularity metrics having systematically bluer energy-shifts (because there is less gravitational red-shifting in the absence of an event horizon), but the main difference between black hole and naked singularity metrics is the appearance of an inner structure for naked singularity image planes which is absent for black hole metrics.  

In naked singularity metrics, photons are observed in a warped image plane structure which is interior to the inner image plane edge of black hole metrics. These photons have passed within $1r_g$ of the singularity at the centre of the Kerr metric, but as there is no event horizon for $|a_\bullet| > 1$, these photons are ``spat out'' from the region around $r\approx 0$  and then make their way to the observer.  The photons in this central cluster originate from the very inner regions of the accretion flow (Fig. \ref{fig:mad-lines}), and so produce a leading order contribution  to observed spectral lines for steep emissivity profiles, like those routinely inferred from spectral fitting \citep[e.g.,][]{Wilms2001,Miller2002,Wilkins2011,Dauser2012}.   This means that even metrics with $a_\bullet = 1 + \delta, (\delta \ll 1)$ can be distinguished from black hole metrics.  For certain observation angles naked singularity metrics produce ``triple'' spectral lines (Fig. \ref{fig:mad-lines}) which would be easily distinguishable from black hole metrics.  

For naked singularities with larger spins $a_\bullet \gtrsim 2$, which have ISCO radii at the same location as black hole metrics with $-0.998 < a_\bullet < 0.998$, it is still possible to distinguish between black hole and naked singularity metrics.  This is trivial for near face-on orientations, as naked singularity metrics produce inverted spectral lines in this limit (Fig. \ref{fig:face-on-inverted-line}), with a sharp red-edge, and extended blue tail. For more edge-on inclinations it is still possible to distinguish the two metrics, providing the emissivity profile of the disc is sufficiently steep ($\gamma \gtrsim 2$; Figs. \ref{fig:different-lines}, \ref{fig:schwarz-like-lines}, \ref{fig:similar-lines}; as is universally required from observations). This is because naked singularity metrics produce broader lines than black hole metrics, owing to larger blue energy shifts (no event horizon) and brighter red wings (larger observed area of red disc regions). This is true for both highly spinning black hole metrics (and their complimentary  naked singularity metrics; Fig. \ref{fig:different-lines}) as well as for the Schwarzschild metric (Fig. \ref{fig:schwarz-like-lines}).

{In this analysis we have assumed that the disc extends down to the ISCO, 
% and thereafter provides 
{with} no significant reflection component
{contributed from inside of the ISCO}.
Of course, in a real accretion system there is non-zero density within the ISCO which could produce a potentially observable spectral component \citep[e.g.,][]{Reynolds13, Wilkins20}. As the density {$\rho$} falls rapidly upon crossing the ISCO \citep[a result of the rapid increase in the radial velocity and fixed radial mass flux e.g.,][]{MummeryBalbus2023}, the disc ionization fraction $\xi \propto 1/\rho$ grows rapidly, and for typical parameters the flow quickly becomes over ionized within the ISCO and this region therefore typically contributes minimally to the time averaged X-ray spectrum \citep{Wilkins20}. It is therefore a poor approximation to treat the intra- and extra-ISCO regions  as having identical rest frame spectra, and for that reason we have neglected these effects in this work. 

This is of course not true for all choices of the free parameters of the theory, and in principle neglecting this region could lead to systematic effects, particularly for naked singularity metrics (which have a significantly larger observable intra-ISCO region). We have computed the energy shift maps (e.g., Fig. \ref{fig:different-images}) for naked singularity metrics including the intra-ISCO region, which are well behaved even in the near-singularity region, and {in future} will include these effects in a more complete treatment which includes radius dependent rest-frame spectra, which is vital in this low density region.    }

We make the code used to produce all of the figures in this paper publicly available as new {\tt XSPEC} models, \skl and \skc. These models take as input 7 parameters, the Kerr metric spin $a_\bullet$, the observer inclination $\theta_{\rm obs}$, the inner and outer disc radii $r_{\rm in/out}$, a break radius $r_{\rm br}$ and two power law emissivity indices $\gamma_{\rm in/out}$. \skl takes as input a line energy $E_{\rm line}$, and computes observed delta-function line profiles such as in Figs. \ref{fig:different-lines}, \ref{fig:similar-lines}, \ref{fig:schwarz-like-lines}, \ref{fig:gamma_dep_lines}, \ref{fig:mad-lines} and \ref{fig:face-on-inverted-line}. The model \skc takes as input a rest-frame X-ray spectrum, and outputs a convolved spectrum like Figs. \ref{fig:convolved}, \ref{fig:convolved_interesting}.  

Coupled with our naked singularity continuum fitting {\tt XSPEC} package \sk \citep{MummeryBalbusIngram24},  all conventional X-ray spectral fitting procedures can now be performed in the fully general Kerr metric, and the cosmic censorship conjecture can be probed with astronomical X-ray observations. We reiterate the philosophy put forward in \citep{MummeryBalbusIngram24}, that the ``measurement'' of spin parameters $a_\bullet > 1$ at high significance may well be telling us of deficiencies in conventional disc modelling. We believe that the inclusion of super-extremal Kerr spacetimes in conventional data analysis pipelines is a win-win addition, as it either increases the confidence we have in conventional approaches (in the result of a rejection of $a_\bullet > 1$ at high significance), or points towards gaps in our understanding of either accretion physics or the nature of compact objects if $a_\bullet > 1$ is required by the data at high significance. 

\section*{Acknowledgments} 
 We would like to thank the reviewer, Chris Reynolds, for a constructive and very helpful review. 
 AM would like to thank  Pedro Ferreira and Georges Obied for stimulating discussions which initiated this work. This work was supported by a Leverhulme Trust International Professorship grant [number LIP-202-014]. For the purpose of Open Access, AM has applied a CC BY public copyright licence to any Author Accepted Manuscript version arising from this submission. AI acknowledges support from the Royal Society. 

\section*{Data availability }
No observational data was used in producing this manuscript. The {\tt XSPEC} models \skl and \skc are available at the following github repository: \url{https://github.com/andymummeryastro/superkerrline}. The continuum fitting {\tt XSPEC} model \sk is available at the following github repository: \url{https://github.com/andymummeryastro/superkerr}

\bibliographystyle{mnras}
\bibliography{andy}

\label{lastpage}

\end{document}